# Online Discoverability and Vulnerabilities of ICS/SCADA Devices in the Netherlands

## Universiteit Twente

Nov
**03**
2020

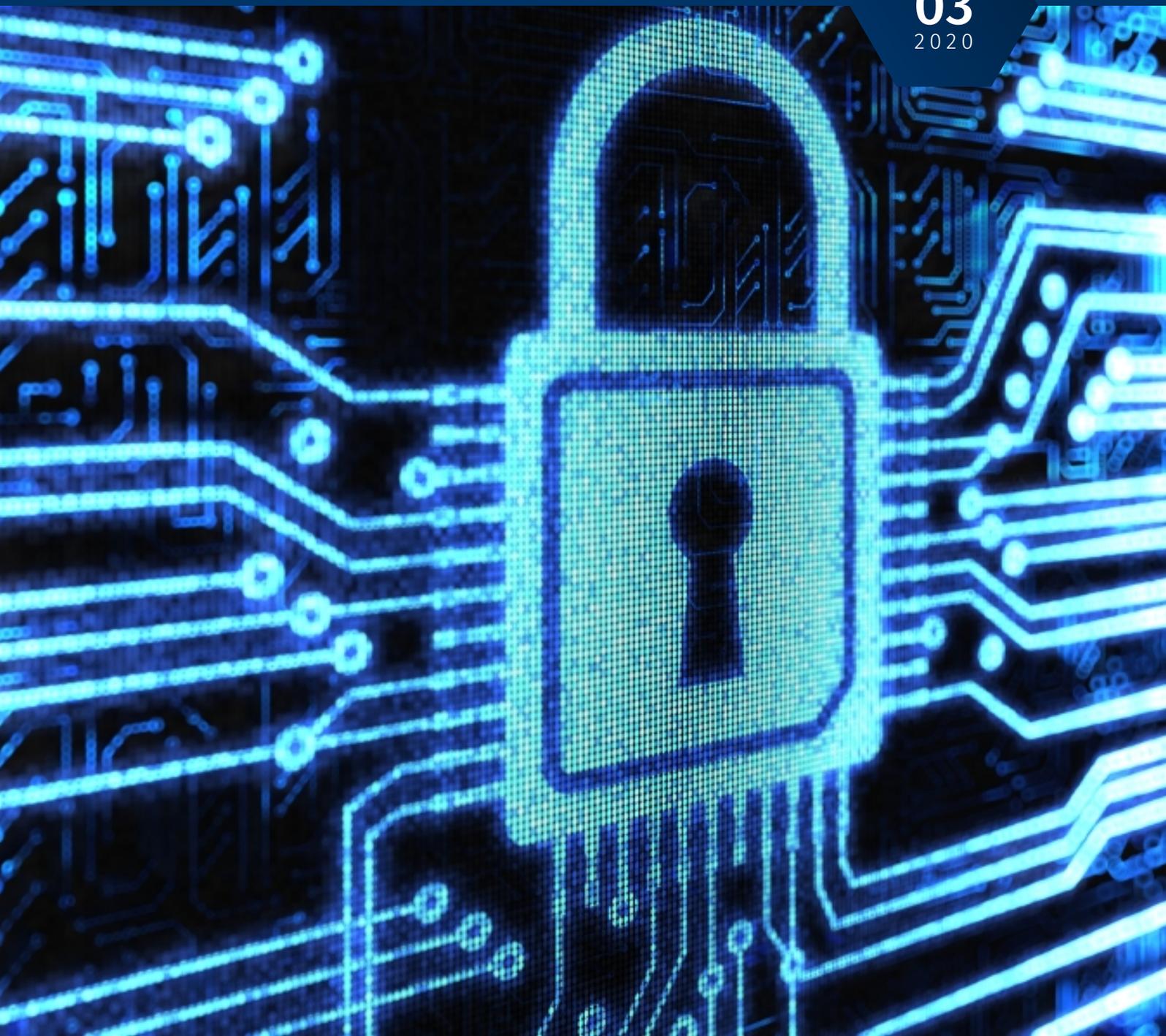

# UNIVERSITY OF TWENTE.

**Authors:**
Dr. J.M. Ceron
Dr. J.J.Chromik
Dr. J.J.C. Santanna
Prof. dr. ir. A. Pras

# Executive Summary

On a regular basis we read in the news about cyber attacks on critical infrastructures, such as power plants. Such infrastructures rely on so-called Industrial Control Systems (ICS) / Supervisory Control And Data Acquisition (SCADA) networks. By hacking the devices in such systems and networks, attackers may take over the control of critical infrastructures, with potentially devastating consequences.

This report focusses on critical infrastructures in the Netherlands and investigates three main questions: *1)* **How many ICS/SCADA devices located in the Netherlands can be easily found** by potential attackers?, *2)* **How many of these devices are vulnerable** to cyber attacks?, and *3)* **What measures** should be taken to prevent these devices from being hacked?

The approach starts with a literature study to determine which ICS/SCADA protocols exist and which TCP/UDP ports are used by these protocols (see Chapter 2). The result of this literature study is a list of 39 protocols, which serves as input to a dedicated search engine (Shodan). The search revealed that, after being queried, almost seventy-thousand systems respond in one way or another. Of these systems only a fraction are real ICS/SCADA devices, the rest are normal PCs, IoT devices etc.. To distinguish between both kind of systems, two lists were created. The first uniquely identifies a system as being an ICS/SCADA device (positive), the second as a non-ICS/SCADA device (negative). In total nearly thousand ICS/SCADA devices were found (see Chapter 3). To determine whether such ICS/SCADA device is prone to known vulnerabilities and to determine the severity of these vulnerabilities, their device signatures were compared to two well known vulnerability datasets (ICS-CERT and NVD, see Chapter 4). Finally, recommendations are provided to limit the discoverability and vulnerability of ICS/SCADA devices (see Chapter 5).

The main conclusions are that *a)* tools like **Shodan** (see Chapter 2) make it extremely easy for potential attackers to find ICS/SCADA devices, *b)* **at least one thousand (989) ICS/SCADA devices in the Netherlands are exposed** on the Internet (see Chapter 3), *c)* around **sixty of these devices have multiple vulnerabilities** with a high severity level (see Chapter 4) and *d)* that several well-known and relatively easy to deploy **measures** exist that help to improve the security of these ICS/SCADA devices (see Chapter 5). .

The goal of this study was to detect vulnerable ICS/SCADA devices in the Netherlands and to propose measures to prevent these devices from being hacked.
At one hand the number of vulnerable devices seems high and worrying, since *in theory the impact* of already a single hacked device may be high (like a lock gate or even power plant failure). In addition, the numbers of 989 and 60 mentioned above must be seen as lower bounds, since this study was limited to only (a) IPv4 addresses, (b) relative straightforward search methods (that can already be used by script kiddies), and (c) well-known vulnerabilities. Professional hackers, such as those working for nation states, are certainly able to find more devices and hack these using zero-day exploits.
On the other hand, this study did not investigate how the detected devices are being used, nor *the real impact* that a hack of one of these devices would have. It is certainly possible that all critical infrastructures in the Netherlands are secure, and that the devices found in this study are not or no longer connected to a critical infrastructure. Therefore we recommend that the results of this study are shared with the critical infrastructure providers, and that further study is performed to better understand the real impact that attacks would have. Finally discussions should start whether it is time to establish a *dedicated Trusted and Resilient network for* *the critical infrastructures* (see also the discussion section at the end of Chapter 6).

#### Samenvatting

Regelmatig verschijnen er nieuwsberichten over cyberaanvallen op vitale infrastructuren, zoals elektriciteit centrales. Dergelijke infrastructuren maken gebruik van zogeheten Industrial Control Systems (ICS) / Supervisory Control And Data Acquisition (SCADA) netwerken. Als dergelijke systemen gehackt worden, kunnen aanvallers de besturing van vitale infrastructuren overnemen, met potentieel enorme gevolgen.

Dit rapport richt zich op vitale infrastructuren in Nederland en beantwoord drie vragen: *1)* **Hoeveel Nederlandse ICS/SCADA systemen zijn eenvoudig te vinden** door potentiële aanvallers?, *2)* **Hoeveel van deze systemen zijn kwetsbaar** voor cyber aanvallen?, en *3)* **Welke maatregelen** kunnen genomen worden om hack pogingen te voorkomen?

Om deze vragen te beantwoorden is bestaande literatuur bestudeerd en uitgezocht welke ICS/SCADA protocollen bestaan, en welke TCP/UDP poorten door deze protocollen worden gebruikt (zie hoofdstuk 2). De uitkomst van deze studie is een lijst met 39 protocollen, die vervolgens gebruikt is als invoer voor een speciale zoekmachine (Shodan). Met behulp van deze zoekmachine zijn bijna zeventigduizend systemen gevonden die één of ander antwoord sturen als ze ondervraagd worden. Hiervan is slechts een klein aantal daadwerkelijke ICS/SCADA systemen, de rest zijn gewone PCs, IoT systemen enz.. Om beide type systemen van elkaar te onderscheiden zijn twee lijsten gemaakt; de eerste zegt met zekerheid of een bepaald systeem een ICS/SCADA systeem is (positief), de tweede met zekerheid dat het geen ICS/SCADA systeem is (negatief). In totaal zijn er bijna duizend ICS/SCADA systemen gevonden die in Nederland op het Internet zijn aangesloten (zie hoofdstuk 3). Om te bepalen welke van deze systemen kwetsbaarheden bevatten, en om de impact van mogelijke aanvallen te bepalen, zijn de kwetsbaarheden vergeleken met bekende lijsten van kwetsbaarheden (ICS-CERT en NVD, zie hoofdstuk 4). Dit rapport eindigt met voorstellen van mogelijke maatregelen waarmee de veiligheid van ICS/SCADA systemen verbeterd kan worden (zie hoofdstuk 5).

De belangrijkste conclusies zijn dat *a)* het met zoekmachines zoals **Shodan** (zie hoofdstuk 2) uiterst eenvoudig is om ICS/SCADA systemen te vinden, *b)* **tenminste duizend (989) Nederlandse ICS/SCADA systemen te vinden zijn** via Internet zoekmachines (zie hoofdstuk 3), *c)* ongeveer **zestig van deze systemen op één of meerdere manieren kwetsbaar zijn** (zie hoofdstuk 4) en *d)* dat er diverse bekende en relatief eenvoudig toepasbare **maatregelen** bestaan waarmee de veiligheid van ICS/SCADA systemen verbeterd kan worden (zie hoofdstuk 5).

Het doel van deze studie was om kwetsbare ICS/SCADA systemen in Nederland te vinden, en maatregelen voor te stellen om te voorkomen dat dergelijke systemen worden gehackt.
Enerzijds lijkt het aantal kwetsbare systemen hoog en reden te geven tot zorg, omdat *in theorie* reeds een enkel gehackt systeem (zoals bijvoorbeeld een sluisdeur of een energiecentrale) grote gevolgen kan hebben. Bovendien is het aantal kwetsbare systemen dat in deze studie genoemd wordt in werkelijkheid waarschijnlijk beduidend hoger, omdat (a) deze studie zich heeft beperkt tot IPv4 adressen, (b) de gebruikte zoekmethode vrij eenvoudig is (en ook door script kiddies toegepast kan worden) en (c) alleen gekeken is naar bekende kwetsbaarheden. Professionele aanvallers, welke bijvoorbeeld voor nationale veiligheidsdiensten werken, zullen zeker meer kwetsbare systemen weten te vinden en in staat zijn binnen te dringen door gebruik te maken van zogeheten zero-day exploits.
Anderzijds is in deze studie niet onderzocht wat de *werkelijke gevolgen* zijn als een systeem wordt gehackt. Het is in principe zeker mogelijk dat alle systemen die zijn aangesloten op de Nederlandse vitale infrastructuren volkomen veilig zijn, en dat de systemen die in deze studie zijn gevonden niet daadwerkelijk worden gebruikt voor vitale diensten.

De belangrijkste aanbevelingen zijn dan ook om de resultaten van deze studie via het NCSC te delen met de organisaties die verantwoordelijk zijn voor de Nederlandse vitale infrastructuur, en verder onderzoek te verrichten teneinde een beter inzicht te krijgen in de werkelijke schade die door aanvallen kunnen worden aangericht. Tenslotte is het tijd om een discussie te starten of er geen *apart veilig en betrouwbaar* netwerk moet komen ten behoeve van de vitale infrastructuren (zie ook de discussie sectie aan het eind van hoofdstuk 6).

# Contents

# 1
Chapter

# Introduction

**Highlights of this chapter:**

- The motivation behind the research described in this report is that various ICS/SCADA devices seem to be inadvertently exposed on the public Internet without proper security measures, potentially causing catastrophic incidents.

- The goals of this report are (1) to quantify how many ICS/SCADA devices located in the Netherlands are easily discoverable and therefore exposed to potential Internet attackers; (2) to quantify the vulnerabilities of these devices; and (3) to provide recommendations for system managers to improve the overall security of these ICS/SCADA systems.

- Our methodology is based on the following steps: (1) collect IP addresses for devices worldwide; (2) filter these IP addresses to find devices located in the Netherlands (NL); (3) classify ICS/SCADA devices among the NL devices; (4) identify the vulnerabilities of NL ICS/SCADA devices based on known vulnerabilities; and (5) identify, as far as possible, the organisations operating these devices.

- From a scientific perspective, our methodology for classifying ICS/SCADA devices extends the state-of-the-art by adding a validation step. This step guarantees that all devices that were classified positively are indeed ICS/SCADA devices. The validation makes use of two lists with signatures. Signatures on the first list identify with certainty devices that are ICS/SCADA devices. Signatures on the second list identify with certainty devices that are *not* ICS/SCADA devices. Both lists were created after extensive analysis of responses from roughly 3 million devices.

*<This page was intentionally left blank.>*

## 1.1 Motivation and Goals

Industrial Control Systems (ICS) are used to monitor and control industrial processes. ICS are usually managed using Supervisory Control and Data Acquisition (SCADA) systems that provide a user interface for operators to monitor and control physical systems. ICS/SCADA devices are used in many sectors, including critical infrastructures, like: (1) power distribution systems, (2) water treatment and sewage facilities, (3) manufacturing facilities, (4) communication facilities, and (5) transportation systems.

Unavailability or failure of critical infrastructures could have serious consequences. Unreliable operation of such systems could disrupt the infrastructure's environment, harm the long-term operation of the organisation responsible for it, or in in the worst scenarios threaten human lives [1]. Examples of large incidents on ICS/SCADA environments include the attacks listed below:

- In December 2015 in the Ukraine hackers (which were likely supported by Russia) left more than two hundred thousand people without electricity by remotely disconnecting several power stations [3];
- One of the biggest aluminum producers in the world, Hydro, was forced to switch to manual operations following a "severe" cyberattack [2];
- In Germany, hackers manipulated and disrupted a steel mill, resulting in massive damage [4];
- In Iran, an attack involving a computer worm, Stuxnet, damaged almost a fifth of the nuclear centrifuges and the damage is estimated to be 1 trillion USD [5].

Note that the examples above are only part of the full picture, as incidents related to critical systems are not often made public. There are also many examples of malware that target ICS/SCADA devices. For instance, the malware called Triton [6], released in 2018, was designed to target a specific product from the manufacturer Schneider Electric. As a consequence, the affected device could be used in spying campaigns by giving control of the device to a remote unauthorised entity.

The General Intelligence and Security Service of the Netherlands (AIVD) has reported the increase of activities that are aimed at facilitating the sabotage of critical infrastructure in Europe [7]. This observation was also noticed by the National Cyber Security Centre (NCSC) of the Netherlands [8]. The NCSC observed that state actors are continuing using digital attacks against other countries. According to the NCSC, significant threats of sabotage and disruption are sponsored by nation-states.

The incidents involving ICS/SCADA systems are a consequence of their evolution. As depicted in Figure 1.1, ICS/SCADA devices systems originally were restricted to being accessed by operators within the infrastructure of the organisation, isolated from the Internet. Service protocols used in these ICS/SCADA devices were therefore designed with functionality as their main goal. It is now desirable for system operators to be able to remotely connect and control the ICS/SCADA systems from anywhere at any time via the Internet [9, 10, 11]. This evolution has several benefits: it facilitates the interoperability of systems and reduces the infrastructure and maintenance costs. However, the ICS/SCADA protocols lack built-in security. Hence, ICS/SCADA devices have been inadvertently exposed on the public Internet without proper security measures, facilitating not only ill-intentioned users (hackers) in gaining access to the devices and potentially causing severe incidents, but also facilitating accidental mistakes by people coincidentally scanning parts of the Internet.

> **The goals** of the research in this report are: (1) to quantify how many ICS/SCADA devices located in the Netherlands are easily discoverable and exposed to any user on the Internet, (2) to quantify the vulnerabilities that these devices have, and (3) to provide recommendations that improve overall security of the exposed ICS/SCADA systems.

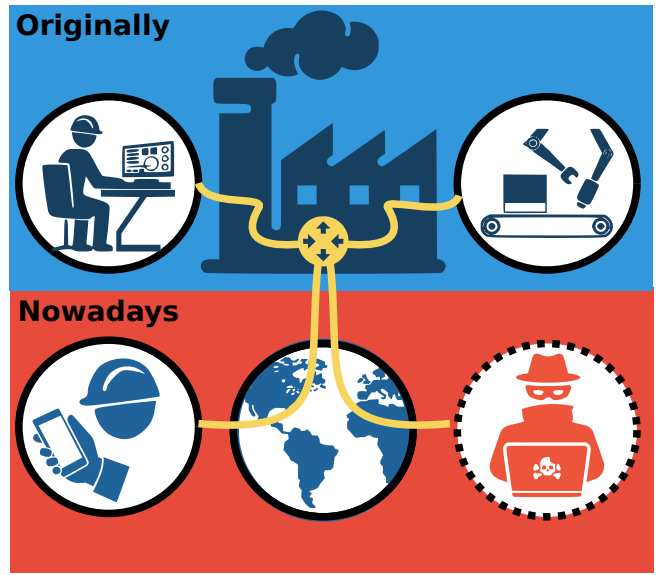

Figure 1.1: Evolution of ICS/SCADA systems.

## 1.2 Concepts and Terminologies

Below we provide some definitions for the terms that are used in the remainder of this report.

> **Vulnerability**: a weakness in the design, implementation or operation of devices that could be exploited to compromise security.

> **Threat**: the danger that emerges once potential vulnerabilities become known and there are people willing and able to exploit that vulnerability.

> **Vulnerable ICS/SCADA device**: a piece of equipment running at least one service with a known vulnerability.

> **Common Vulnerabilities and Exposures (CVE)**: is a reference-method for publicly known information-security vulnerabilities and exposures. A CVE is usually related to versions of services running on a device.

> **Public IP address**: is an IP address that can be accessed over the Internet.

> **Network Address Translation (NAT)**: a method of remapping private IP address(es) into public IP address(es) and vice-versa. NAT, for example, allows several home user devices to access the Internet with a single public IP address.

> **Autonomous System (AS)**: a collection of Internet addresses, controlled by a network operator, that

> share the same routing policies. An AS is identified by its number (ASN) and its name (AS_name).

> **Internet Service Provider (ISP)**: an organisation that provides services for accessing the Internet. Every ISP has one or more ASes, but not all ASes are ISPs.

> **ICS/SCADA device / product / system**: an ICS/SCADA device is a piece of hardware that performs one or many ICS/SCADA services. We use the term product when we want to associate a manufacturer to a device. Note that also other concepts, such as Distributed Control Systems (DCS) and Building Automation Systems (BAS) exist, which are comparable to ICS/SCADA systems. For the purpose of this report these concepts will be considered to be equivalent.

> **ICS/SCADA protocol and service**: ICS/SCADA protocol is a communication language shared between ICS/SCADA devices on a specific port number. We use the term service to refer the implementation of a communication protocol.

> **ICS/SCADA port number**: a number smaller than 65.535 that identifies "where" a protocol/service is running on the ICS/SCADA device.

To clarify the concepts and terminologies, we present an example in Table 1.1, retrieved from `https://www.shodan.io/host/130.89.14.205`. The example shows one desktop machine, with a single public IP address, which is managed by the UTWENTE AS (number 1133). This single device has three ports open: 22, 80, and 443, which are used by the protocols SSH, HTTP, and HTTPS, respectively. These three protocols deploy in the OpenSSH and Apache httpd. Note that two protocols (HTTP and HTTPS) point to a single service, Apache httpd. Finally, nine vulnerabilities, indicated with their CVE numbers, are known for this version of the Apache httpd service.

Table 1.1: Example of device information retrieved from `https://www.shodan.io/host/130.89.14.205` for clarifying the terminology used in this report.

| IP Address | ASN | AS name | Device | Ports | Protocols | Services | Vulnerabilities |
|---|---|---|---|---|---|---|---|
| 130.89.14.205 | AS1133 | UTWENTE | <not available> | 22 | SSH | OpenSSH | - |
| | | | | 80 | HTTP | Apache httpd | CVE-2018-1302<br>CVE-2017-15710<br>CVE-2018-1301<br>CVE-2018-1283<br>CVE-2018-1303<br>CVE-2017-15715<br>CVE-2018-1333<br>CVE-2018-11763<br>CVE-2018-1312 |
| | | | | 443 | HTTPS | Apache httpd | |

## 1.3 Report Structure and Overall Methodology

This report is structured into five parts (see Figure 1.2). In Chapter 2 we provide the background, which is essential for understanding the later chapters. In Chapter 3 we describe how to find in the Netherlands those ICS/SCADA devices that are accessible to any Internet user. In Chapter 4 we explain how to discover the vulnerabilities within the identified devices'. In Chapter 5 we provide recommendations to improve the overall security of the exposed ICS/SCADA systems. Chapter 6 provides the conclusions.

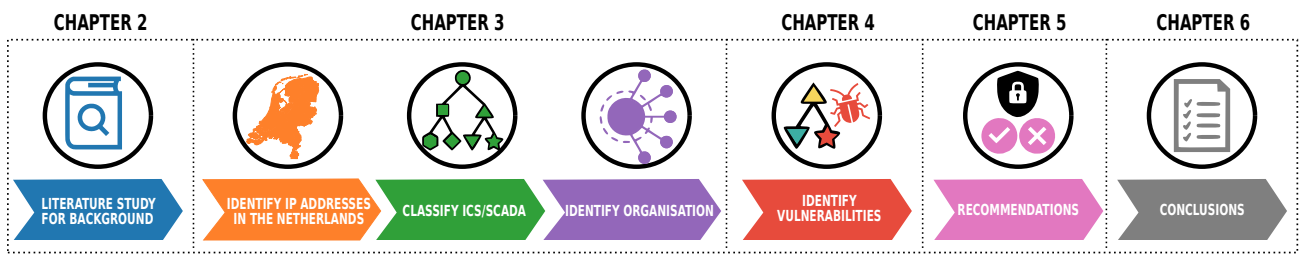

Figure 1.2: Overall document structure and methodology.

Our methodology starts with a literature study to determine (1) the best known ICS/SCADA protocols and ports, (2) the best known tools to scan devices, and (3) the best known projects that port scan all devices connected to the Internet.

The second part of this report, in which we determine the discoverability of ICS/SCADA devices in the Netherlands, uses a threefold methodology. First, we retrieve all IP addresses geolocated in the Netherlands. Second, these IP addresses are classified as either ICS/SCADA devices or not. Third, the organisations that connect these ICS/SCADA devices to the Internet are identified.

The third part of this report, in which we investigate the vulnerabilities of ICS/SCADA devices in the Netherlands, compares the characteristics of ICS/SCADA devices to a list of known vulnerabilities.

The last part of this report provides recommendations to improve the security of ICS/SCADA devices , followed by the conclusions and discussion how severe the results are.

## 1.4 Scope of the Report and Target Audience

This section highlights some aspects that are **not** covered by the research in this report and aspects that could create technical, legal, or ethical issues. In addition, we describe the target audience:

- For the methodology in Chapter 3 we do **not** port scan IP addresses ourselves. The reason for this is that the act of scanning creates potential technical, legal, and ethical issues. For example, some ICS/SCADA devices reboot when some types of scans are performed. To circumvent these issues, we decided to use information from the Shodan project, which carefully port scanned a comprehensive set of devices (using IP version 4 addresses – IPv4) in the Netherlands. The implication of this decision is that our results are dependent on the correctness of the dataset provided by the Shodan project. More details are given in Chapter 3.

- For the methodology in Chapter 3, we do **not** investigate devices connected to the Internet via IP address version 6 (IPv6). To the best of our knowledge, there is **no open** project that provides this type of information. Brute-force scanning of the IPv6 address space is not possible. For example, currently it is possible to scan the $2^{32}$ (*i.e.*, more than four billion) IPv4 addresses in a couple of hours, however IPv6 has $2^{128}$ valid addresses (*i.e.*, 340.282.366.920.938.463.463.374.607.431.768.211.456), which would be too much to scan (not considering the impact of the volume of requests generated). To overcome this limitation, we could have investigated the relation between IPv4 and IPv6 address. However, this aspect

- is out of the scope for the research within this report. The implication of our decision to not investigate IPv6 devices is that the number of devices that we found may be lower than the actual number of devices.

- For the methodology in Chapter 3, we decided to use a list with *default* port numbers of the *most widely-known* devices and protocols in ICS/SCADA (in Table 2.1). Therefore this methodology does **not** identify known protocols running on ports other than the default. Our methodology is also restricted to the list that we collected on 'most widely-known ICS/SCADA protocols and ports'. The implication of this decision is that the number of devices that we found may be lower than the actual number of devices.

- The methodology in Chapter 4, to identify the known vulnerabilities of ICS/SCADA devices is mainly based on information provided by a North American organisation (ICS-CERT). The reason for this choice is that this organisation provides the most comprehensive database of vulnerabilities in the world. The implication of this choice is that it potentially contains more information on North American manufactures/products. Unfortunately, we were unable to find another comprehensive dataset focusing on European manufacturers. This fact does not explicitly affect the findings in this report, as manufacturers of ICS/SCADA devices are mostly international. However, it is possible that not all vulnerable devices were discovered.

**The target audience** of this report are policy makers, ICS/SCADA experts working at vendors and critical infrastructure providers, as well as security experts working at ISPs and organisations such as the National Cyber Security Centre (NCSC).

*<This page was intentionally left blank.>*

# 2
Chapter

# ISC/SCADA Device Discoverability

**Highlights of this chapter:**

- The goal of this chapter is to explain the methodology used to find potential ICS/SCADA devices.

- Based on literature study, we developed a list with the 39 best-known ICS/SCADA protocols and port numbers.

- After comparing the best-known Internet scanning projects, we concluded that the Shodan project provided the best results for the purpose of this study.

- We observed that, to classify an ICS/SCADA devices , we need (1) the value of the port number and (2) meta-data returned by the device.

*<This page was intentionally left blank.>*

## *2.1* Goal and Chapter Structure

The goal of this chapter is to explain essential concepts to understand our methodology, which we will present in the next chapter. This chapter is divided into two parts: (1) determining the most common ICS/SCADA protocols and port numbers and (2) strategies to discover generic devices connected to the Internet.

## *2.2* ICS/SCADA Protocols and Port Numbers

It is important for this study to collect a comprehensive list of protocols and the default port numbers used by ICS/SCADA devices. This will be the first type of information needed to find potential ICS/SCADA devices. Our methodology is based on a literature search, using the top ten most cited academic papers retrieved by Google Scholar using the keywords "ics scada scan". We chose to use the most cited papers because we consider these the most relevant material related to ICS/SCADA devices [12, 13, 14, 15, 16, 17, 18, 19, 20, 21]. In addition to the top 10 most relevant academic papers, we used the protocols and ports related to ICS/SCADA, as listed by Censys and Shodan. Table 2.1 shows our list.

Table 2.1 identifies 39 ICS/SCADA protocols. There are protocols that share the same port number. For example, the protocols ICCP (line 15), IEC 61850 MMS (line 17), and Siemens S7 (line 34) operate by default on port 102; Danfoss ECL apex (line 8) and SAIA S-BUS (line 32) on port 5050; ProConOS (line 30) and Schleicher XCX 300 (line 33) on port 20547; and EtherNet/IP (line 10) and YASKAWA MP2300Siec (line 38) on port 44818. Besides those running on the same ports, there are also protocols that use multiple ports. For example, EtherNet/IP (line 10) runs on port 2222 and 44818, GE-SRTP (line 12) on port 18245 and 18246, LS Fenet (line 20) on port 2004 and 2005, MELSEC Q (line 21) on port 5006 and 5007, and Unitronics Socket1 (line 36) on port 20256 and 20257. This happens because these protocols use different transport protocols (TCP and UDP).

Based on the findings that (1) single port numbers can lead to multiple protocols and (2) single protocols can operate using multiple ports, we conclude that a methodology based only on port numbers is not sufficient for classifying ICS/SCADA. Hence, we opted to enhance our methodology by using the meta-data information provided by the services running on the ICS/SCADA devices . In the following section, we describe in more detail the meta-data used to classify a ICS/SCADA devices and also the approach used to search for devices in the Netherlands.

Table 2.1: Well known ICS/SCADA protocols and ports.

|   | Protocol | Default Port |
|---|---|---|
| 1 | ANSI C12.22 | 1153 |
| 2 | BACNet | 47808 |
| 3 | Beckhoff-ADS communication | 48898 |
| 4 | CANopen | 7234 |
| 5 | CodeSys | 2455 |
| 6 | Crimson 3 | 789 |
| 7 | DNP3 | 20000 |
| 8 | Danfoss ECL apex | 5050 |
| 9 | EtherCAT | 34980 |
| 10 | EtherNet/IP | 44818,2222 |
| 11 | FATEK FB Series | 500 |
| 12 | GE-SRTP | 18245,18246 |
| 13 | HART-IP | 5094 |
| 14 | HITACHI EHV Series | 3004 |
| 15 | ICCP | 102 |
| 16 | IEC 60870-5-104 | 2404 |
| 17 | IEC 61850 / MMS | 102 |
| 18 | KEYENCE KV-5000 | 8501 |
| 19 | KOYO Ethernet | 28784 |
| 20 | LS Fenet | 2005,2004 |
| 21 | MELSEC Q | 5006,5007 |
| 22 | Modbus/TCP | 502 |
| 23 | Moxa | 4800 |
| 24 | Niagara Tridium Fox | 1911,4911 |
| 25 | OMRON FINS | 9600 |
| 26 | OPC | 135 |
| 27 | PCWorx | 1962 |
| 28 | Panasonic FP (Ethernet) | 9094 |
| 29 | Panasonic FP2 (Ethernet) | 8500 |
| 30 | ProConOS | 20547 |
| 31 | Quick Panel GE | 57176 |
| 32 | SAIA S-BUS (Ethernet) | 5050 |
| 33 | Schleicher XCX 300 | 20547 |
| 34 | Siemens S7 | 102 |
| 35 | Simatic | 161 |
| 36 | Unitronics Socket1 | 20256,20257 |
| 37 | YASKAWA MP Series Ethernet | 10000 |
| 38 | YASKAWA MP2300Siec | 44818 |
| 39 | Yokogawa FA-M3 (Ethernet) | 12289 |

## *2.3* ICS/SCADA Devices Discoverability

As previously described, this report is an investigation of devices reachable over the Internet. There are over 4 billion IP version 4 (IPv4) address. Therefore, to identify the devices located in the Netherlands, automated tools are required. In this section we describe the most widely-known automated tools and projects that use these tools for detecting devices. We conclude this chapter by identifying which project is able to find the highest number of devices located in the Netherlands, which will be used in the remainder of this report.

Port scan is the act of checking whether a port number of a device is open or closed (a further explanation about port numbers is given in § 1.2). Although these types of tools are extremely useful for network operators to discover and monitor the status of devices running in a network, ill-intentioned users also use them for reconnaissance and misuse of devices. There are several tools that perform port scans, five examples are presented in Table 2.2.

Table 2.2: Port Scanning tools.

|   | Scanning Tool | Reference |
|---|---|---|
| 1 | Nmap | Lyon [22] |
| 2 | Zmap | Durumeric et al. [23] |
| 3 | Masscan | Graham [24] |
| 4 | Unicornscan | Louis [25] |
| 5 | Dscan | Song [26] |

The first port scan tool in Table 2.2 is one of the oldest and most widely known in the security community, Nmap. This tool was first released in 1997 and using it makes it possible to scan the entire IPv4 address space in a couple of months. Zmap, the second tool listed in Table 2.2 , was released in 2013 and it is able to run 1,300 times faster than the Nmap (less than one hour for the entire IPv4 address space, but only for a single port).

For the research in this report we decided to **not** use any port scanning tool ourselves. The reason is that there are already several projects that port scan the entire IPv4 address space on a daily basis. We do **not** want to generate unnecessary network traffic against any device. In addition, there is an ethical/legal discussion on whether active measurements such as port scan should be performed [27]. For example, when the Heartbleed vulnerability was discovered in 2014, many researchers started to scan for vulnerable systems. An unintended side effect of these scanning attempts was that several systems crashed.
For this study we therefore decided to rely on existing and known projects that already perform port scans. These projects are accepted by the security community and overcome some of the ethical/legal issues.

## Port Scanning Projects

Following the evolution of port scan tools, several projects have emerged. There are several projects that use port scan tools for explicitly scanning the entire Internet. Five examples of projects are presented in Table 2.3. In this table, the two most well-known projects are Shodan and Censys. The first, Shodan, advertises itself as "the world's first search engine for Internet-connected devices". Since 2013, this project port has been scanning the entire IPv4 address space and updating their database in real-time. Shodan is a private initiative and does not reveal the port scan tool used. Censys was created in 2015 at the University of Michigan, by the researchers who developed the Zmap port scan tool. Over the past four years, the team has performed thousands of Internet-wide scans, consisting of trillions of probes, and has played a central role in the discovery or analysis of some of the most significant Internet-scale vulnerabilities, such as FREAK, Logjam, DROWN, Heartbleed, and the Mirai botnet. At the end of 2018, Censys turned the project into a private initiative.

Table 2.3: Port Scanning Projects and Respective Scanning Tool.

|   | Scanning Project | Scanning Tool | Reference |
|---|---|---|---|
| 1 | shodan.io | - | Shodan [28] |
| 2 | censys.io | ZMap and ZGrab | Censys [29] |
| 3 | zoomeye.org | Xmap and Wmap | KnownSec [30] |
| 4 | rapid7.com | - | Rapid7 [31] |
| 5 | kudelskisecurity.com | - | Kudelski Security [32] |

It is important to highlight that any project that performs an Internet-wide scan generates a large amount of network traffic. As a consequence, these tools can affect the normal operation of some devices, particularly legacy devices, such as most of the old ICS/SCADA devices . Besides these scanning projects, malicious software (malware) within infecting machines (related to botnets) also performs Internet-wide scans in their reconnaissance phase, for example the Mirai botnet. There are several initiatives for monitoring Internet-wide scans, such as the one by Morris [33] and the one by the Center for Applied Internet Data Analysis (CAIDA) [34]. This type of monitoring initiative is important for identifying the origin of port scans, which can be used to block malicious types of network activity. Although this monitoring project could contribute to blocking malicious activities, this is out of the scope of the research in this report.

For the research in this report, it is important to define which port scanning project is able to identify the most comprehensive list of devices in the Netherlands, particularly those related to ICS/SCADA devices (more details on our methodology in § 3.1). In our preliminary analysis, we investigate what is collected by each scanning project (1) and did a literature study (2) to determine the most common protocols. Despite the list of common protocols associated with ICS/SCADA devices (presented in Table 2.1), we have observed that scanning projects only consider a subset of them. The project Censys, for example, only evaluate 4 protocols related to ICS/SCADA devices. On the other hand, Shodan has a broader view in terms of ICS/SCADA devices by evaluating 16 protocols. Table 2.4 presents the protocols' comparison of its coverage by the Shodan and Censys search engines (for a detailed description see Appendix A).

Table 2.4: SCADA protocols supported by the most popular device search engines. Source: [35] and [13].

| Engine | BACnet | CodeSys | Crimson v3 | DNP3 | EtherNet/IP | GE-SRTP | HART-IP | IEC60870-5-104 | IEC61850 | MELSEC-Q | Modbus | Tridium | OMRON-FINS | PCWorx | ProConOS | Siemens S7 |
|---|---|---|---|---|---|---|---|---|---|---|---|---|---|---|---|---|
| shodan.io | ✓ | ✓ | ✓ | ✓ | ✓ | ✓ | ✓ | ✓ | ✓ | ✓ | ✓ | ✓ | ✓ | ✓ | ✓ | ✓ |
| censys.io | ✓ | | | ✓ | | | | | | | ✓ | | | | | ✓ |

As shown in the table, Shodan has better coverage of ICS/SCADA devices protocols. Moreover, the outcome of our queries to both databases was that Shodan returned more IP addresses geolocated in the Netherlands, because Shodan queries more port numbers than Censys. This finding contradicts to the results in [13], which is a paper written by researchers from Censys. Our preliminary investigation of both search engines also revealed that, while Censys takes a snapshot of all IPv4 address devices every single day, Shodan takes around two weeks to query the entire Internet. The reason is that Shodan splits the scanning to cover more ports than Censys and gathers additional information about the devices. Both projects geolocate IP addresses, but they do not declare which database is used. Examples of databases are [36, 37, 38, 39].

The final finding is that both projects include meta-data retrieved from devices (banner). This meta-data is, in general, a configurable "welcome" text from the scanned device. This meta-data usually provides system information, *e.g.*, data about the operating system (OS), software/firmware versions, and web services running in a specific port number. When a device is not configured, it displays default information, which can include sensitive information or access to login screens. If configured, a banner can have a custom message set by the administrator, which could be (i) obfuscating the information about the service, or (ii) providing misinformation to confuse malicious parties. Sometimes a banner can provide an unreadable response, if a service cannot process the request properly. In this report, the banner information is essential for validating which devices are ICS/SCADA devices, and which devices are *not* ICS/SCADA devices (a further explanation is provided in § 3.1).

# 3
Chapter

# Exposed ICS/SCADA Devices in the Netherlands

**Highlights of this chapter:**

- In the Netherlands, 3,09 million devices are connected to the Internet. Almost one thousand (989) of these devices could be classified as ICS/SCADA devices. This number is substantial, considering that **anyone** connected to the Internet is able to access those devices.

- The Tridium manufacturer could be related to more than five hundred ICS/SCADA devices (557), which represents 55% of all ICS/SCADA devices in the country. One explanation for this large number of Tridium devices is their generic nature, which makes it possible to use them in any sector.

- Most of the ICS/SCADA products are used to enable legacy ICS/SCADA equipment to connect to the Internet. Alarmingly, we observe that these devices do not have built-in security. We therefore advise managers and operators of ICS/SCADA systems to replace legacy equipment with more secure equipment.

- We investigated both physical and cyber locations of the ICS/SCADA devices. This localisation relied on Autonomous System (AS) numbers. We discovered that only the Internet Service Providers to which critical infrastructure providers are connected could be identified. Apparently critical infrastructure organisations rely for routing and protection on general ISP services.

## *3.1* Methodology to Classify ICS/SCADA Devices

Our methodology to classify ICS/SCADA devices is based on multiple steps.

First, we collect from the Shodan project all IP addresses located in the Netherlands. As explained in the previous chapter, Shodan is a search engine that takes two weeks to scan the entire Internet. Our dataset was retrieved in an incremental way on a daily basis between 28 May and 19 June 2018. From the collected data we removed the duplicate entries that have the same IP address and port number.

Second, we filter the IP addresses that are associated to ICS/SCADA ports/protocols. For this filtering we use the list with 39 ICS/SCADA protocols and ports as described in the previous chapter (Table 2.1). Note that, if one IP address has at the same time an ICS/SCADA port (from our list) as well as other generic ports (outside our list), then we took all ports for further analysis. The reason for this is to detect also devices that are intentionally configured to use port numbers different from the default.

Third, for each IP address (and port number) identified in the previous step, we analysed the meta-data that was retrieved by Shodan. Meta-data is basically the 'welcome' message that is returned by the device after a connection request has been received. We compare the content of the meta-data to a list of positive and negative features (Appendix B). These features are strings or keywords that tell if a service is *de facto* related to ICS/SCADA or not.

Examples of positive features related to port 102 are: "Siemens", "61850", "SIMATIC", "6ES7", and "TS_600_GOLD". This means that, if an IP address has port 102 open and returns, as part of the connection establishment phase, meta-data that contains one of these keywords, the device can be positively classified as an ICS/SCADA device. Similarly, examples of negative features are: "FTP", "SSH", "Conpot"(type of honeypot), and Deathmatch (game server). Thus if an IP address returns one of these words as part of its meta-data, it will be classified as *not* being an ICS/SCADA device. The list of features was created after analysing more than three million generic devices (IP addresses) located in the Netherlands. The total list with positive and negative features used in this research is available in Appendix B. The meta-data of IP addresses that does not match with any feature (positive or negative) is labelled as 'not-classified'.

## *3.2* Findings

### *3.2.1* Overall number of ICS/SCADA Devices

In Figure 3.1, we summarise our overall findings. Using Shodan, we found that 3,09 million IP systems are located in the Netherlands. These systems can be reached via more than one thousand (1.220) Autonomous Systems (ASes). On average, each device is running 1,9 service (5,98 million services in total). Of these 3,09 million devices, 68.166 devices (2.2%) had an open port that potentially relates to an ICS/SCADA protocol. Since one device can have multiple open ports (services), we found in total 71.816 services running on these systems.

After running our classification methodology (see § 3.1), we found that there are almost one thousand (989) ICS/SCADA devices in the Netherlands. This number of devices represents only 0,02% of all devices in the country (3.09 million). This percentage is somehow within the range we could expect, since ICS/SCADA devices are relatively special kind of devices. We also observed that the average number of active services that run on an ICS/SCADA device is 1,2 (1215 services in total). This finding reinforces that ICS/SCADA devices are primarily used for single applications. The 989 ICS/SCADA devices can be related to only sixty (60) products from twenty five (25) manufacturers. The devices are reachable via 85 Autonomous Systems (ASes)

In the remaining of this chapter we will discuss the most common products, manufacturers and the organisations that operate ICS/SCADA devices.

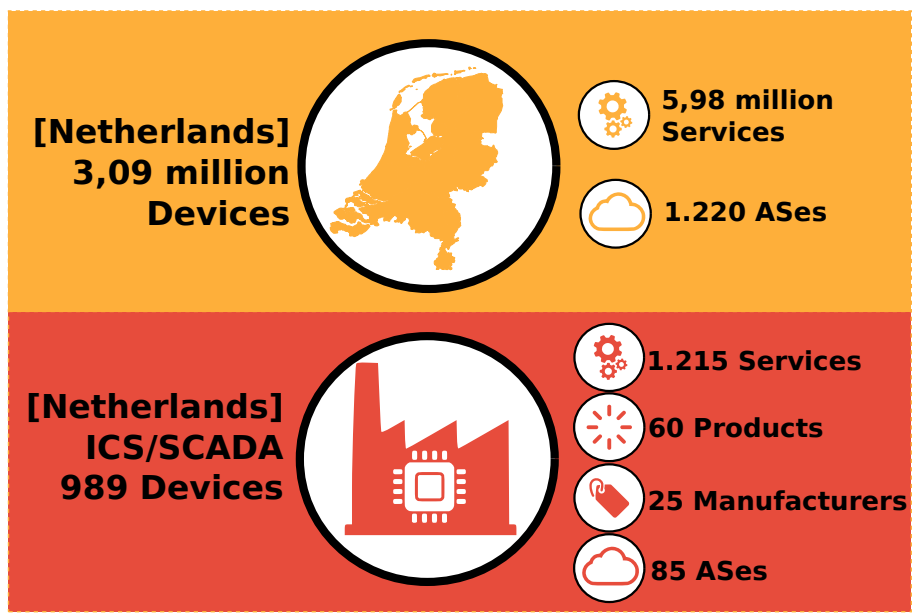

Figure 3.1: Publicly accessible devices in the world and in the Netherlands.

## 3.2.2 Manufacturers Related to ICS/SCADA Devices

In Figure 3.2 we show the top 10 manufacturers of ICS/SCADA devices within the Netherlands. The complete list of manufacturers (25) can be found in Appendix D.

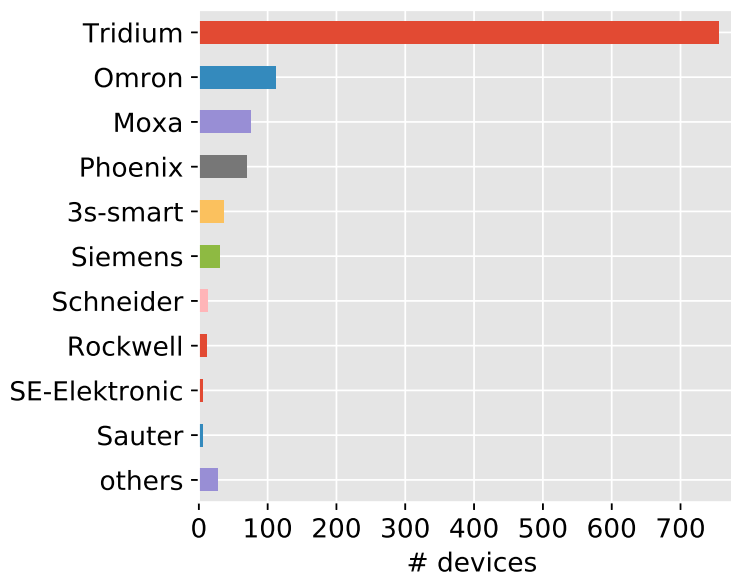

Figure 3.2: Top 10 manufacturers of ICS/SCADA devices in the Netherlands.

Tridium is responsible for more than five hundred ICS/SCADA devices in the Netherlands (557 devices). This number represents more than fifty percent the number of ICS/SCADA devices in the country (55,31%). Tridium is an American company founded in 1995; it makes products that enable the integration of building automation and other engineering control systems (*e.g.*, Modbus, DeviceNet, EtherNet/IP, CANopen, PROFIBUS and PROFINET networks). Their main products enable legacy protocols to interoperate with a single control system. This integration capability could be one reason on why we found so many devices from this manufacturer in the Netherlands.

Omron accounts for five times less devices than Tridium (112 devices). This Japanese company was founded in 1933 and builds automation components, equipment and systems. Although this company is generally known for medical equipment (*e.g.,* digital thermometers, blood pressure monitors and nebulizers), the second position may also be related to the functionality provided by these devices, which is enabling legacy devices to be managed in a single manner.

Phoenix and Moxa have a very similar number of devices, 69 and 67, respectively. While the former was founded in 1923 in Germany, the latter was founded in 1987, in the United States. The following companies 3s-smart, Siemens, Schneider, Rockwell, SE-Elektronic, and Sauter account for less than 5% of the devices in the Netherlands.

The most important conclusion is that Tridium is responsible for the highest number of discoverable devices in the Netherlands.

### 3.2.3 ICS/SCADA Products

In Figure 3.3 we show the top 10 most common ICS/SCADA products in the Netherlands. The complete list with 60 ICS/SCADA products can be found in Appendix C.

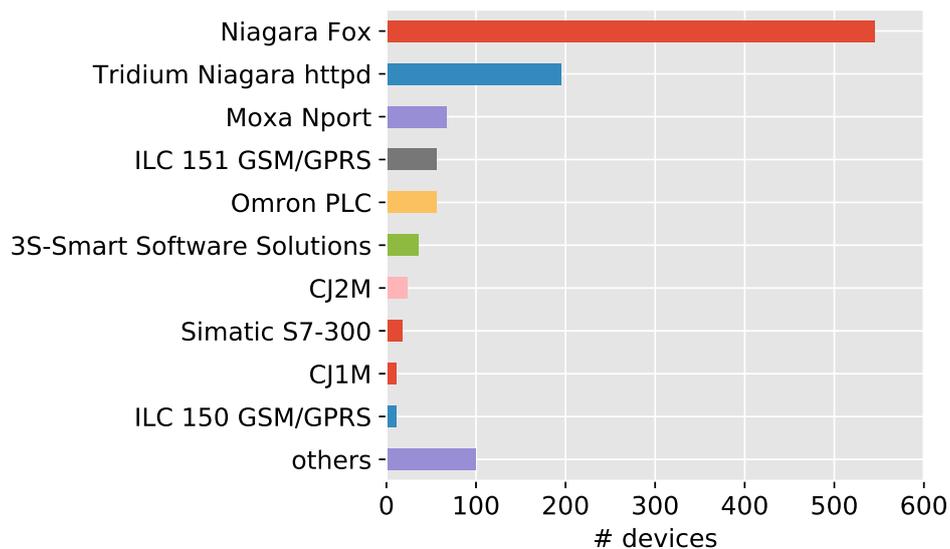

Figure 3.3: Top 10 ICS/SCADA devices type in the Netherlands.

As expected, the top 10 products are mainly coming from the top 10 manufacturers (Figure 3.2). For example, Niagara Fox (545 devices) and Tridium Niagara httpd (195 devices) are both products from Tridium. However, the order in which the top products appear is not the same as the order in which the top manufacturers appear. For example Omron, which is the top 2 manufacturer, occupies the 5th position with the Omron PLC, the 7th position with the CJ2M and the 10th position with the CJ1M. Another example is Phoenix, which occupies the 4th position as manufacturer and occupies the 4th and 9th position with the ILC 151 GSM/GPRS and ILC 150 GSM/GPRS.

### 3.2.4 Organisations Operating ICS/SCADA Devices

In the ideal case we would be able to map the IP addresses of the discovered devices to the organisations that operate these devices. However, according to the GDPR, IP addresses should be considered as personal data and are therefore privacy sensitive (this discussion is in fact a bit more subtle, and gives different outcomes for the US and UK, countries that are traditionally less privacy sensitive. However, such discussion is outside the scope of this report). Lists that show the mapping between IP addresses and the organisations that use these IP addresses are therefore not publicly available. Although ISPs would be able to create such lists, sharing such lists with researchers would (most likely) be illegal.

Instead of mapping *individual* IP addresses to organisations, it is possible however to map *sets of related* IP addresses to organisations. Such sets are created for routing purposes; all addresses within the same set shares the same routes to and from systems elsewhere on the Internet. Such sets of Internet addresses are called Autonomous Systems (AS).

In Figure 3.4 we show the top 10 Autonomous Systems related to ICS/SCADA devices located within the Netherlands. That figure is based on information obtained from Shodan [28], and enriched with AS specific information obtained from Team Cymru [40]. The complete list with all 85 ASes can be found in Appendix E.

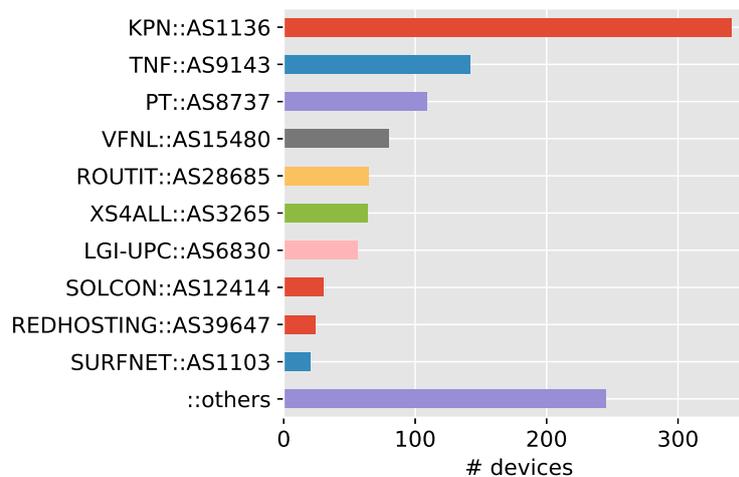

Figure 3.4: Top 10 ASes with ICS/SCADA devices in the Netherlands.

The most interesting finding is that the top 10 ASes are all ISPs. This means that none of the ASes points directly to an actual ICS/SCADA organisation. This means that ICS/SCADA organisations are 'protected' behind (and thus dependent on) their ISPs. After a manual analysis we discovered that the top 1 (KPN) and top 3 (PT) belongs to KPN. We also observed that the top 2, top 4, and top 7 belongs to Liberty Global (that was before Vodafone and Ziggo). This observation means that ICS/SCADA infrastructures are connected to the Internet via the main telecommunication companies.

Security by obscurity is never sufficient, however. If one of these ISPs becomes victim of a large Distributed Denial of Service [DDoS] attack, then all ICS/SCADA devices within that ISP may loose connectivity. Therefore we recommend to start the discussion whether a dedicated *Trusted and Reliable* network for critical infrastructures should be established.

An interesting finding is that SURFNET (AS1103), the academic ISP, occupies the 10th position, the University of Eindhoven (AS1161) occupies the 47th position, and the University of Twente (AS1133) occupies the 79th position (more ASes in Appendix E). This shows that the academic community is investigating (the security of) ICS/SCADA devices. Let's hope they will bring improvements to the security level of the society.

Finally, although Shodan [28] provides the geolocation (latitude and longitude) of devices, in the majority of cases this information is misleading. As we explained in this section, the IP address of ICS/SCADA devices can be related to ISPs, and not to ICS/SCADA organisations. Therefore, the geolocation information provided by Shodan points to the routers and headquarters of ISP, and not the device. For example, imagine a device located in Enschede connected to the Internet via KPN ISP. In this example the location of the device will be Amsterdam, and not Enschede.

# 4
Chapter

# ICS/SCADA Devices Vulnerabilities in the Netherlands

**Highlights of this chapter:**

- Our approach to classify whether ICS/SCADA devices are vulnerable or not uses three pieces of meta-data collected from the ICS/SCADA devices: manufacturer, service, and service version. These pieces of meta-data are compared to two publicly available list of vulnerabilities: ICS-CERT and NVD. In addition to the vulnerability classification, we propose a methodology to assess the severity of vulnerabilities, based on the well-known CVSS method.

- Of the 989 ICS/SCADA devices found in the previous section, only 6% (63) had one or more vulnerabilities. However, we expect that most of these devices can be easily exploited by hackers, with possibly unforeseeable consequences.

- We found 37 distinct vulnerabilities. All devices have at least one vulnerability with a high level of severity. This means that all 63 vulnerable devices have a critical need to be patched, which would (maybe surprisingly) be easy to do.

- The vulnerable devices come from only five vendors: Omron, Siemens, Rockwell, Schneider, and Tridium. This does not mean that these vendors built 'insecure' products, but that organisations that use ICS/SCADA devices from these manufacturers seems to be reluctant to patch these devices.

*<This page was intentionally left blank.>*

# *4.1* Methodology For Classifying Device Vulnerability

Our methodology to identify vulnerabilities of ICS/SCADA devices relies on comparing the meta-data information collected from the ICS/SCADA devices (provided by Shodan) with information from a publicly available list of vulnerabilities. We use three characteristics within the meta-data from ICS/SCADA devices: (1) the manufacturer of the device, (2) the service running in the device, and (3) the version of this service. We use two well-known databases with lists of vulnerabilities: one from the Industrial Control Systems Cyber Emergency Response Team (ICS-CERT) [41] and one from the National Vulnerability Database (NVD) [42]. Usually when a vulnerability is identified, the security community proposes an update to the service (software) version running on the device. Therefore, the threefold meta-data information is sufficient for identifying whether an ICS/SCADA device remains vulnerable (using the same version of the service) or not.

Another important aspect of the ICS/SCADA meta-data information is that the data is non-structured (a set of strings). It required effort to extract the threefold information from the meta-data. Some information in the meta-data was too vague to reveal the service of ICS/SCADA systems and for some devices the meta-data information was blank. In the first case, to retrieve the threefold set of information, we manually accessed Websites from manufacturers and security teams, such as Talos [43] and Siemens [44]. For the second case, blank meta-data, we did *not* perform the vulnerability classification.

After extracting the threefold meta-data information, we compared it with two well-known databases, ICS-CERT and NVD. Although our analyses are only based on these two sources of information, to the best of our knowledge these are the sources with the most comprehensive lists of vulnerabilities. An implication of using only these sources of information is that the number of vulnerable ICS/SCADA devices found in this chapter is potentially lower than the actual number. Upon request, we can make the script with the analysis available to researchers who would like to extend our analysis by including other sources of information.

In addition to identifying the vulnerabilities, we had the initial intention to assess the *risk* of a vulnerability to a company. However, performing this type of assessment requires a considerable amount of information related to the organisation, such as the type of organisation, how critical the service provided by the organisation is, where the device is placed, and what the function of the device is in the organisation's infrastructure. For example, a vulnerable device controlling the energy facility for an entire city is usually considered more risky than if such device is used to control an energy facility for a single user (such as a solar panel used in a residence). These aspects are out of the scope of the research in this report. Therefore, instead of a risk assessment we decided to assess the severity of vulnerabilities.

For assessing vulnerability severity we use a method proposed by the National Institute of Standards and Technology (NIST) [45], which assigns ranges of scores into three severity levels: low (from 0 to 3.9), medium (from 4 to 6.9) and high (from 7 to 10). The scoring method is called Common Vulnerability Scoring System (CVSS) and it provides a vulnerability score number between 0 and 10. It takes into account many features, such as the type of attack vector (local/remote), the attack complexity, the privileges required, the user interaction, the scope, and how the device security is affected in terms of confidentiality, integrity, and availability.

Both datasets of vulnerabilities that we used in the methodology of this chapter (ICS-CERT and NVD) already provide either the CVSS or the severity level for the Common Vulnerabilities and Exposures (CVE) listed in their database. Therefore, for the analysis in this chapter, for each vulnerability found we also collected or calculated the severity level (low, medium, or high). There are some limitation related to CVSS, such as those highlighted by McAfee Labs [46]. These limitations do *not* invalidate their value in this chapter. Once again, the implication of these limitations is that the number of vulnerable ICS/SCADA devices found in this chapter is potentially lower that the actual number.

## *4.2* Findings

First, we consider all the ICS/SCADA devices in the Netherlands that we found in the previous chapter of this report as being vulnerable, thus devices that can easily be reached via the Internet. In this section, we investigate specific types of vulnerabilities, known by the security community as Common Vulnerabilities and Exposures (CVE). This section is divided into (1) the overall findings on the vulnerability of ICS/SCADA devices in the Netherlands, (2) a detailed explanation of the CVEs found in the Netherlands, (3) a severity analysis of each CVE, (4) an analysis of vulnerabilities by manufacturer, (5) product type, and (6) organisation.

### *4.2.1* Overall ICS/SCADA Vulnerabilities

Based on the methodology described in § 4.1, in Figure 4.1 we highlight the overall vulnerabilities found in ICS/SCADA devices in the Netherlands.

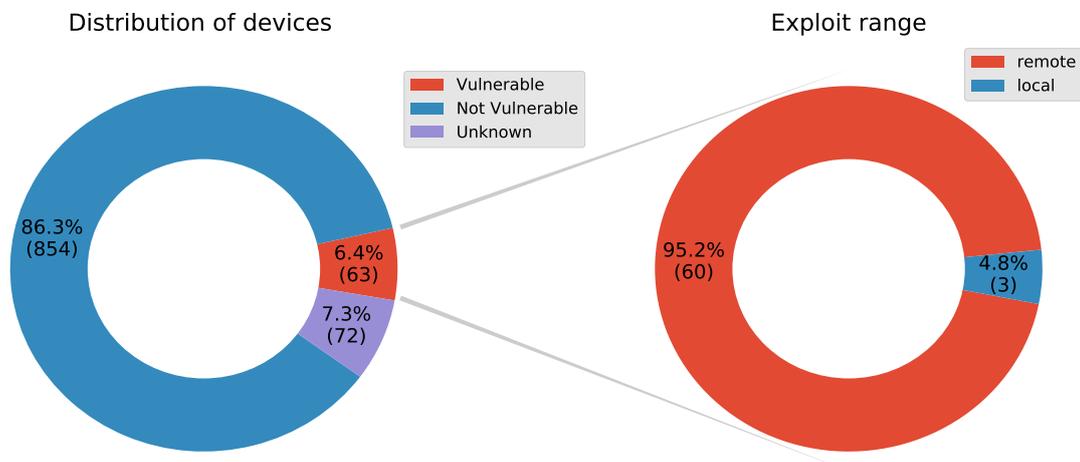

Figure 4.1: Percentage of devices that are vulnerable and possibility to exploit the vulnerability remotely.

In Figure 4.1, the graph on the left shows the percentage of devices that are vulnerable; the graph on the right shows whether a vulnerability can be exploited remotely. The graph on the left shows that, from the 989 ICS/-SCADA devices found in the previous chapter, 63 devices show one or more vulnerabilities (6% of the total). Although this 6% suggests that most organisations that operate ICS/SCADA devices have fixed vulnerabilities, still a large number of devices can be exploited by any ill-intentioned user (hackers) connected to the Internet. Exploiting in this case implies, for example, collecting sensitive information from the organisation, making the device inaccessible, executing any type of remote code, and bypassing service authentication. Depending on where the device is placed and the type of organisation, compromising the device could cause catastrophic incidents.
72 devices (7%) are classified as *unknown*, since we did not have enough information to classify them.
Of the 63 vulnerable devices ICS/SCADA devices (graph on the right), we observed that 95% (60) can be exploited remotely, meaning that any hacker can exploit the device remotely, thus without the need to physically access to the device. 5% of the devices (3 in total) require local access to be exploited.

Figure 4.2 shows that the majority of the devices have two or more vulnerabilities. For example, 28 devices have two vulnerabilities and 21 devices have five vulnerabilities.

Although on average each device has 5 vulnerabilities, this value is not representative because it is not normally distributed. We were therefore surprised to find eight devices with 16 vulnerabilities. There are several plausible explanations for this finding. For example, the owners of those devices may not be sufficiently aware of security best practices. However, it may also be that these devices are somehow forgotten, since they are no longer used to control critical infrastructures. Finally it may even be that these devices are used as honeypots, intended to attract and identify attackers.

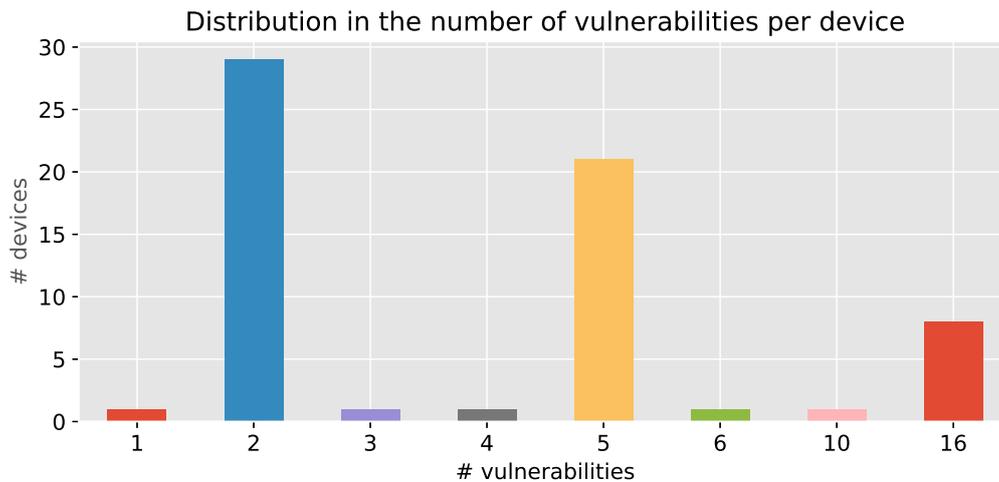

Figure 4.2: This figure shows the distribution in the number of vulnerabilities per device.

## 4.2.2 Specific Vulnerabilities of ICS/SCADA Devices

After investigating the overall vulnerabilities of ICS/SCADA devices in the Netherlands, in this section we describe the specific vulnerabilities (CVEs) that we found. In Table 4.1, we present the vulnerabilities, the manufacturer of the vulnerable device, the type of exploration range, the severity level, the number of occurrences, and the number of unique devices with this vulnerability. We sort the content of the table based on the number of occurrences of a vulnerability.

In Table 4.1, there are 37 unique vulnerabilities in total. 34 of these vulnerabilities have a high level of severity, which is critical to the device and, consequently, to the organisation. More investigation and discussion on this aspect can be found in § 4.2.3. Five vulnerabilities have a number of occurrences different from the number of unique devices (highlighted in bold text). For example, the vulnerability `CVE-2017-2680` appears 24 times in 23 devices. The reason is that there are devices running two identical services (in different ports) with the same vulnerability. To provide a better understanding of the vulnerabilities, we briefly describe them, indicating whether there is a solution or mitigation for the problem.

- **CVE-2015-0987** – affects the specific version of products (CJ2M, CJ2H, and CX-Programmer) from the manufacturer Omron. For this vulnerability (sensitive) account information is transmitted without encryption. An unauthorised user could intercept this sensitive information and compromise the device remotely. Solution/Mitigation: manufacturer released a software update;

- **CVE-2015-1015** – targets multiple products (CJ2M/CJ2H) from the manufacturer Omron. This vulnerability enables an unauthorised user to read sensitive information from the device. Solution/Mitigation: manufacturer released a software update;

- **CVE-2017-2681** – this vulnerability affects multiples products from the manufacturer Siemens, which uses the protocol PROFINET. Successful exploitation of this vulnerability could cause the targeted device to enter a denial-of-service condition, which may require human interaction to recover the system. Solution/Mitigation: the manufacturer has released a software update for a subset of products;

- **CVE-2017-2680** – this vulnerability is related to CVE-2017-2681. Again, a specially crafted packet can be used to cause the target device to enter into a state that may require human intervention for recovery. This CVE identification affects another subset of products (SIMATIC HMI Multi and S7-300/S7- 400). Solution/Mitigation: manufacturer released a software update;

- **CVE-2017-12741** – affects multiple products from the manufacturer Siemens, including the products Sinamics/SIMATIC/SIMOTION. When exploited, this vulnerability can turn the device inaccessible. An unauthorised user can, over the Internet, crash the SCADA device by denying its services to legitimate users. Solution/Mitigation: manufacturer released a software update;

- **CVE-2015-2177** – affects all versions of the product SIMATIC S7-300 from the manufacturer Siemens. This vulnerability allows the performance of a denial of service (DoS) attack over the network without prior authentication. A cold restart is required to recover the system. Specially crafted packets sent to Port 102/TCP can be used to stop the device and demand a restart. Solution/Mitigation: the manufacturer does not provide a specific software update to solve the problem, however, it proposed the use of migration methods to avoid device exposure, such as VPN and access restriction;

- **CVE-2016-9158** – affects all the families of the product SIMATIC S7-300 and SIMATIC S7-400 from the manufacturer Siemens. Successful exploitation of this vulnerability means the device needs to be restarted to recover the system. Solution/Mitigation: manufacturer released a software update;

- **CVE-2016-9159** – also affects all the families of the product SIMATIC S7-300 and SIMATIC S7-400 from the manufacturer Siemens, targeting the protocol ISO-TSAP and Profibus. Successful exploitation of this vulnerability enables an unauthorised user to get sensitive information including the device credentials. Solution/Mitigation: manufacturer released a software update;

- **CVE-2017-14462, CVE-2017-14463, CVE-2017-14464, CVE-2017-14465, CVE-2017-14466, CVE-2017-14467, CVE-2017-14468, CVE-2017-14469, CVE-2017-14470, CVE-2017-14471, CVE-2017-14472, CVE-2017-14473, CVE-2017-12090, CVE-2017-12089, CVE-2017-12088** – this set of vulnerabilities is associated with the product Micrologix 1400 Series B FRN from the manufacturer Rockwell. A specially crafted packet can cause a read or write operation resulting in disclosure of sensitive information, modification of settings, or modification of the sequential logic (ladder logic). These vulnerabilities can be exploited remotely and do not require any authentication to trigger them. Solution/Mitigation: manufacturer has released a software update;

- **CVE-2017-16740** – affects a specific version of the product MicroLogix 1400 Controllers from the manufacturer Rockwell. Successful exploitation of this vulnerability could cause the device to become unresponsive to Modbus TCP communications and affect the availability of the device. Solution/Mitigation: manufacturer has released a software update;

- **CVE-2017-6030, CVE-2018-7789, CVE-2018-7790, CVE-2018-7791, CVE-2018-7792** – these vulnerabilities affect multiple versions of the product Modicon from the manufacturer Schneider. Successfully exploiting these flaw allows unauthorised users to obtain sensitive information, reboot the system, upload files, and overwrite the password. Solution/Mitigation: manufacturer has released a software update;

- **CVE-2017-16744** – this vulnerability affects multiple versions of the product Niagara from the manufacturer Tridium. When successfully exploited, an unauthorised user can obtain administrator credentials. Solution/Mitigation: manufacturer has released a software update;

- **CVE-2012-4701** – this vulnerability affects multiple versions of the product Niagara from the manufacturer Tridium. This flaw enables unauthorised users to read sensitive files and execute arbitrary code. Solution/Mitigation: manufacturer has released a software update;

- **CVE-2012-4027, CVE-2012-4028** – these vulnerabilities affect multiple versions of the product Niagara AX Framework from the manufacturer Tridium. When successfully exploited, an unauthorised user can read the configuration file and bypass access restrictions. Solution/Mitigation: manufacturer has released a software update;

- **CVE-2012-3024, CVE-2012-3025** – these vulnerabilities affect a specific version of the product Niagara AX Framework from the manufacturer Tridium. An unauthorised user can exploit cryptographic flaws to bypass the authentication process via brute-force attacks. Solution/Mitigation: manufacturer has released a software update;

- **CVE-2015-7937** – this vulnerability affects the Modicon M340 product line from the manufacturer Schneider. When successfully exploited, an unauthorised user can execute arbitrary code remotely on the device. Solution/Mitigation: manufacturer has released a software update;

- **CVE-2016-7090** – this vulnerability affects the product SCALANCE from the manufacturer Siemens. Exploitation of this vulnerability could allow an unauthorised user to get access to sensitive information. Solution/Mitigation: manufacturer has released a software update.

Table 4.1: List of vulnerabilities found on ICS/SCADA devices in the Netherlands.

|    | Vulnerability  | Manufacturer | Type   | Score | Severity | Occurrences | Unique Devices |
|----|----------------|--------------|--------|-------|----------|-------------|----------------|
| 1  | CVE-2015-0987  | Omron        | Remote | 10.0  | high     | 25          | 25             |
| 2  | CVE-2015-1015  | Omron        | Local  | 2.1   | low      | 25          | 25             |
| 3  | CVE-2017-2680  | Siemens      | Local  | 6.1   | medium   | **24**      | **23**         |
| 4  | CVE-2017-12741 | Siemens      | Remote | 7.8   | high     | **20**      | **19**         |
| 5  | CVE-2015-2177  | Siemens      | Remote | 7.8   | high     | **19**      | **18**         |
| 6  | CVE-2016-9158  | Siemens      | Remote | 7.8   | high     | **19**      | **18**         |
| 7  | CVE-2016-9159  | Siemens      | Remote | 8.6   | high     | **19**      | **18**         |
| 8  | CVE-2017-14464 | Rockwell     | Remote | 10.0  | high     | 8           | 8              |
| 9  | CVE-2017-14473 | Rockwell     | Remote | 10.0  | high     | 8           | 8              |
| 10 | CVE-2017-14472 | Rockwell     | Remote | 10.0  | high     | 8           | 8              |
| 11 | CVE-2017-14471 | Rockwell     | Remote | 10.0  | high     | 8           | 8              |
| 12 | CVE-2017-14470 | Rockwell     | Remote | 10.0  | high     | 8           | 8              |
| 13 | CVE-2017-14469 | Rockwell     | Remote | 10.0  | high     | 8           | 8              |
| 14 | CVE-2017-14468 | Rockwell     | Remote | 10.0  | high     | 8           | 8              |
| 15 | CVE-2017-14467 | Rockwell     | Remote | 10.0  | high     | 8           | 8              |
| 16 | CVE-2017-14466 | Rockwell     | Remote | 10.0  | high     | 8           | 8              |
| 17 | CVE-2017-14465 | Rockwell     | Remote | 10.0  | high     | 8           | 8              |
| 18 | CVE-2017-14463 | Rockwell     | Remote | 10.0  | high     | 8           | 8              |
| 19 | CVE-2017-14462 | Rockwell     | Remote | 10.0  | high     | 8           | 8              |
| 20 | CVE-2017-12090 | Rockwell     | Remote | 7.8   | high     | 8           | 8              |
| 21 | CVE-2017-12089 | Rockwell     | Remote | 7.8   | high     | 8           | 8              |
| 22 | CVE-2017-12088 | Rockwell     | Remote | 7.8   | high     | 8           | 8              |
| 23 | CVE-2017-16740 | Rockwell     | Remote | 10.0  | high     | 8           | 8              |
| 24 | CVE-2017-2681  | Siemens      | Local  | 6.1   | medium   | 4           | 4              |
| 25 | CVE-2017-6030  | Schneider    | Remote | 10.0  | high     | 4           | 4              |
| 26 | CVE-2018-7789  | Schneider    | Remote | 7.8   | high     | 4           | 4              |
| 27 | CVE-2018-7790  | Schneider    | Remote | 10.0  | high     | 4           | 4              |
| 28 | CVE-2018-7791  | Schneider    | Remote | 10.0  | high     | 4           | 4              |
| 29 | CVE-2018-7792  | Schneider    | Remote | 10.0  | high     | 4           | 4              |
| 30 | CVE-2017-16744 | Tridium      | Remote | 8.0   | high     | 2           | 2              |
| 31 | CVE-2012-4701  | Tridium      | Remote | 9.3   | high     | 2           | 2              |
| 32 | CVE-2012-4028  | Tridium      | Remote | 10.0  | high     | 2           | 2              |
| 33 | CVE-2012-4027  | Tridium      | Remote | 10.0  | high     | 2           | 2              |
| 34 | CVE-2012-3025  | Tridium      | Remote | 10.0  | high     | 1           | 1              |
| 35 | CVE-2015-7937  | Schneider    | Remote | 10.0  | high     | 1           | 1              |
| 36 | CVE-2016-7090  | Siemens      | Remote | 10.0  | high     | 1           | 1              |
| 37 | CVE-2012-3024  | Tridium      | Remote | 10.0  | high     | 1           | 1              |

For some vulnerabilities is possible to find an exploit (software designed to take advantage of a flaw in a system, typically for malicious purposes) [47]. This means, that an attacker does not have to develop tools to explore the vulnerable devices, making the process easier. Some flaws are very simple to exploit, for example the vulnerabilities CVE-2017-12088. To exploit this vulnerability an attacker could send a simple packet to the service running on the port 44818/TCP, as illustrated in the code below:

```
echo -e "\x00\x00\xE8\xFF\x00\x00\x00\x00\x00\x00\x00\x00\x00\x00\x00\x00\x00\x00\x00\x00\x00\x00\x00\x00"
    | nc -w 2 <target_IP> 44818 > /dev/null
```

This code could affect 8 devices in the Netherlands. This means that an attacker with non-advanced skill can compromise 8 ICS/SCADA devices and possibly affecting critical infrastructure.

Another interesting point is the ageing of the vulnerabilities. The CVE code includes the year the vulnerability was reported, for example, the CVE-2018-7789 date the year 2018. The majority of the vulnerabilities found were reported in 2017, however it is possible to observe vulnerabilities from 2012, such as CVE-2012-4028, CVE-2012-4701, CVE-2012-4027, CVE-2012-3025, CVE-2012-3024. This suggests that those devices (8 in total) have been vulnerable for more than 7 years.

It is essential to observe that all the vulnerabilities found already have ways to fix or mitigate these vulnerabilities. Fixing usually involves performing a software update. Based on our findings, we may conclude that these vulnerable devices either have negligent security or were not updated due to organisational policies. We understand that some companies prefer to manage the risk and avoid possible instability caused by a software update. However, as described in this chapter, the majority of the vulnerabilities found can remotely be exploited via the Internet, and since they can be found by using search engines such as Shodan, they can be easily compromised by attackers.

### 4.2.3 ICS/SCADA Vulnerability Severity Level

In this section we discuss the level of severity of vulnerabilities found in ICS/SCADA devices. As described at § 4.1 our methodology is based on values of the open standard CVSS. In Figure 4.3 we show our observations.

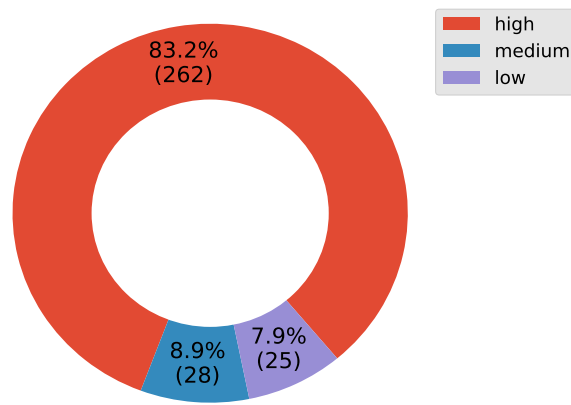

Figure 4.3: Distribution of vulnerability severity on the discovered devices.

In Figure 4.3, 83.2% of the vulnerabilities are classified with high severity, 8.9% as medium, and 7.9% as low severity. It is important to note that previously we found that ICS/SCADA devices frequently have more than one vulnerability (Figure 4.2). After an extensive analysis we observed that all 63 devices (including three devices that can only be exploited with local access) have at least one vulnerability with a high level of severity. This means that all devices are *extremely vulnerable* to being compromised by any ill-intentioned user on the Internet.

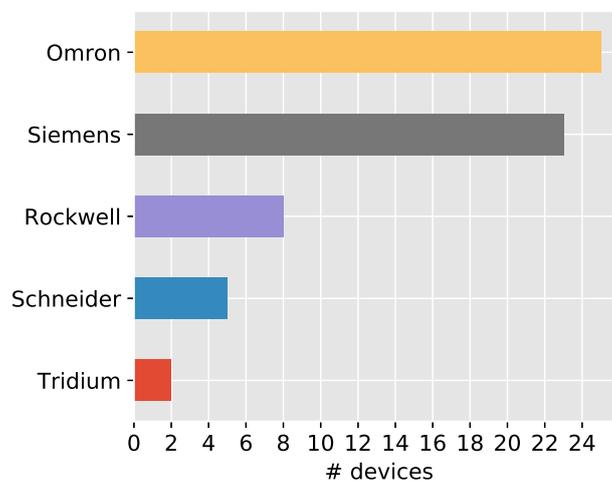

Figure 4.4: The number of vulnerable devices per manufacturer.

## 4.2.4 ICS/SCADA Vulnerabilities by Manufacturer

The goal of this section is to show the landscape of vulnerabilities per manufacturer, with the aim of identifying any relation between vulnerable devices and manufacturers. First, in Figure 4.4, we show the number of vulnerable devices per manufacturer. Of the 25 manufacturers related to ICS/SCADA devices in the Netherlands (discussed at § 3.2.2), only five manufacturers are related to vulnerable devices. As depicted in the figure, we observe that the manufacturer Omron has the highest number of devices with vulnerabilities, followed by Siemens, Rockwell, Schneider, and Tridium.

In Figure 4.5a we present the number of vulnerable devices considering the top 10 general manufacturers of ICS/SCADA devices in the Netherlands. We observed that the five manufacturers related to vulnerable devices are among the top 10 general manufacturers. We also observe that some of these top 10 manufactures have **no** vulnerable device, for example Moxa, 3S-Smart, SE-Elektronic, and Sauter. Note that in Figure 4.5a we can barely see the 2 vulnerable devices from the manufacturer Tridium. We are surprised to observe that almost all ICS/SCADA devices from Siemens are vulnerable. This finding does not mean that Siemens is an insecure manufacturer but that organisations that use Siemens ICS/SCADA devices are reluctant to deploy modifications for improving the security. In Figure 4.5b, we observe that there is no overall relation between vulnerable devices and not vulnerable devices from the same manufacturer.

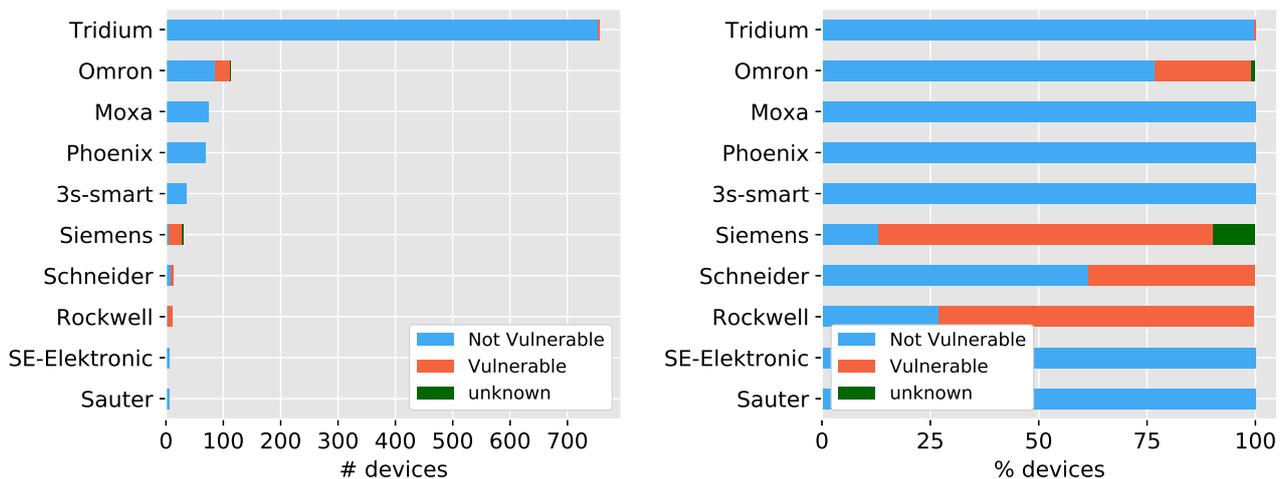

(a) The number of devices per manufacturer.  (b) Percentage of devices per manufacturer.

Figure 4.5: Top 10 NL ICS/SCADA Vulnerable and Not Vulnerable manufacturers.

## *4.2.5* ICS/SCADA Vulnerabilities by Product

The goal of this section is to investigate the most vulnerable ICS/SCADA products. Figure 4.6 shows all the vulnerable ICS/SCADA products we found and their respective manufacturers. Out of the 63 total ICS/SCADA vulnerable devices, we have found only 13 distinct products from the five manufacturers (Figure 4.5a). The vulnerable product with most occurrences in the Netherlands is the `Omron CJ2M`, which was found in 23 devices, followed by Simatic S7-300 (18 devices) from Siemens, and MicroLogix from Rockwell (7 devices). The manufacturer Omron has two products listed (CJ2M and CJ2H), however as seen in Figure 4.5b the majority of the products found are not vulnerable. To understand the types of vulnerable devices, we describe the characteristics of the top three most popular ones in the Netherlands.

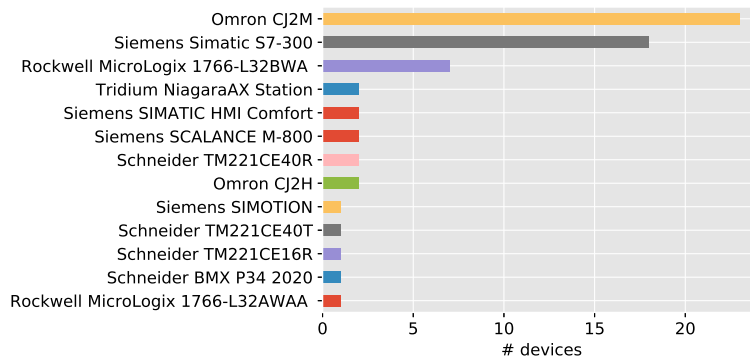

Figure 4.6: ICS/SCADA vulnerable products.

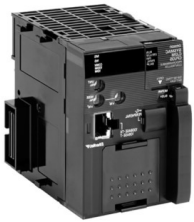

Figure 4.7: CJ2M.

The device **CJ2M** (Figure 4.7), from manufacturer Omron, is the most vulnerable in the Netherlands. CJ2M is a multi-purpose device used in automation. The device has a built-in USB port and the choice of Ethernet and RS-232C/422/485 interfaces on the CPU. Additional modules could be added including a power supply; digital and analogue input and output; motion/position Control Units; and other communication network interfaces.

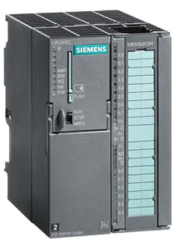

Figure 4.8: SIMATIC S7-300.

The **SIMATIC S7-300** (Figure 4.8) is a general purpose device used in industrial automation. As with the CJ2M, many modules could be added to expand the functionality of the device. Besides communication modules, it is possible to find solutions applied for several sectors including chemical, critical manufacturing, dams, defence industrial base, energy, food and agriculture, government facilities, transportation systems, and water and wastewater systems. We investigate in which sector this device is actually used in the next section.

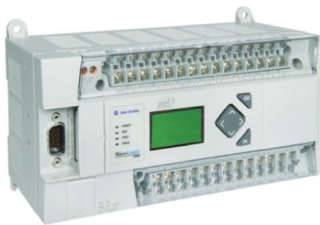

Figure 4.9: MicroLogix 1766-L32BWA.

The **MicroLogix 1766-L32BWA** is a programmable controller from the manufacturer Rockwell and is the third most vulnerable in the Netherlands (Figure 4.9). This product is used for multi-purpose industrial automation and it supports the protocol EtherNet/IP, Modbus TCP/IP, and DNP3 over IP. The device supports expansion modules that provide more flexibility in terms of communication capability, input, and output signal processing and could also be monitored using a Web interface.

## *4.2.6* ICS/SCADA Vulnerabilities by Organisation

The goal of this section is to identify where the vulnerable ICS/SCADA devices are located in the Netherlands, by organisation and sector. Similar to in § 3.2.4, we were unable to meet this goal. The reason is that the IP address of ICS/SCADA devices are currently pointing to Autonomous Systems Numbers (ASN) from Internet Service Providers (ISP). This means that organisations running ICS/SCADA devices either do not run their devices in their own AS or do not have a dedicated infrastructure to forward their traffic. Moreover, it is not clear how the network routing policy is performed by the organisations and how vulnerable they are in terms of routing misconfiguration and abuses. Figure 4.10 presents our findings.

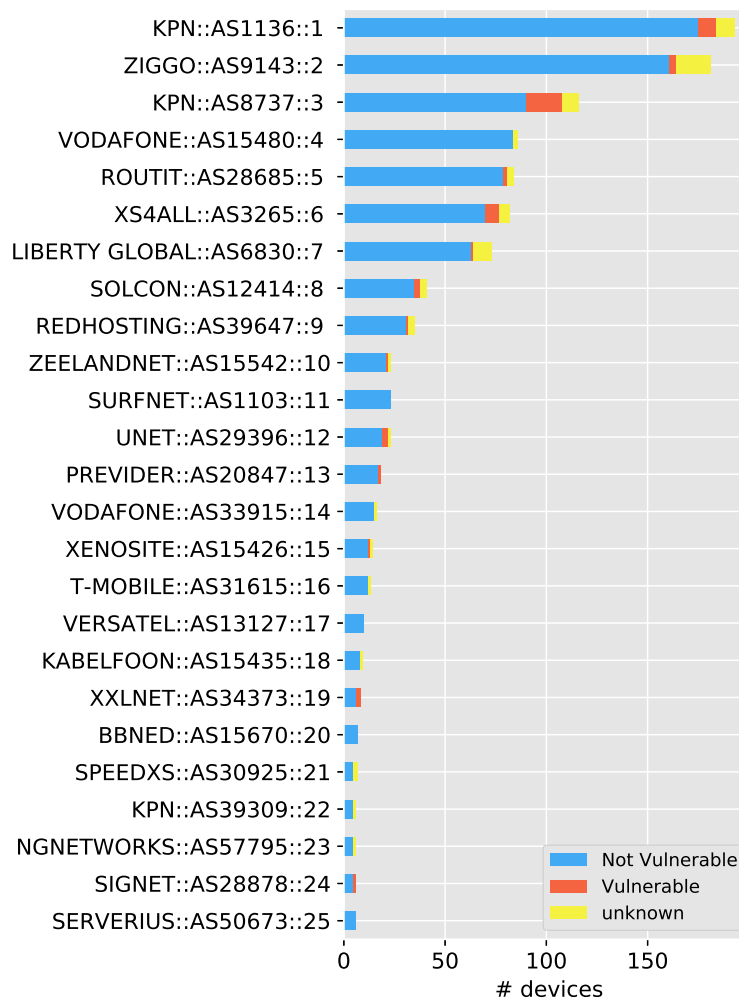

Figure 4.10: Distribution of vulnerable devices by ASes.

Although there are vulnerable devices in 85 ASes, we show only the top 25 in Figure 4.10. The reason is that the other ASes have only a couple of vulnerable devices (less than 6). All these 25 ASes are associated with ISPs and transit providers. Note that although some ASes are repeated, their numbers are different meaning different networks. For example, KPN, position 1, 3, and 22, is related to AS1136, AS8737, and AS39309.

The most important finding is that the biggest ISPs in the Netherlands are the ones that host most of the vulnerable devices. For example, KPN (position 1, 3, and 22), and the ISPs ZIGGO (position 2) and VODAFONE (position 4 and 14) are part of LIBERTY GLOBAL (position 7). Similar to the discussion at § 3.2.4 we conclude in this section that on the one hand, the organisations effectively hide themselves behind ISPs but on the other hand the ISPs become critical. If an ISP such as KPN goes offline due to an attack, for example a Distributed Denial of Service attack, then all the organisations behind KPN will be severely affected.

# 5
Chapter

# Measures to be taken

**Highlights of this chapter:**

- The goal of this chapter is to propose measures and recommendations to improve the security of ICS/SCADA devices in the Netherlands.

- Most measures are well-known, easy to implement and in-line with earlier factsheets issued by the National Cyber Security Centre.

- To prevent the discovery of ICS/SCADA devices using search engines such as Shodan, ICS/SCADA devices should no longer be directly accessible from the open Internet, but hidden behind firewalls, Virtual Private Networks (VPNs) and/or Virtual Local Area Networks (VLANs).

- Techniques should be applied that restrict network traffic on ports and protocols associated with ICS/SCADA services. Examples of such techniques are rate-limiting and the whitelisting of legitimate users. Restriction not only provides protection against potential hacking attempts, but also against Denial-of-Service (DoS) and/or brute force attacks.

- It is recommended that a discussion gets initiated whether a *dedicated Trusted and Resilient network for the critical infrastructures* should be established. To a certain extent such dedicated network would resemble the existing 112 network. Current ISPs could work together to establish such network.

*<This page was intentionally left blank.>*

## 5.1 Measures

Vulnerabilities on ICS/SCADA devices pose a significant threat to industrial networks, particularly those associated with critical infrastructures. To improve the protection of these infrastructures, we recommend a set of measures to secure these devices and to reduce the chance that attacks are successful. These measures are basically well-known, easy to implement and in-line with earlier factsheets issued by the National Cyber Security Centre, such as the 'Checklist security of ICS/SCADA systems' NCSC [48].

- Limit the access of ICS/SCADA devices from the Internet. This can be accomplished by, for example the use of firewalls, Virtual Private Networks (VPNs) or Virtual Local Area Networks (VLANs). Only devices that *must* have external communication may have a direct connection to the Internet.

- Install software updates in a timely manner. When it is not feasible to update the software, make sure the device can not be accessed via the Internet.

- To avoid that ICS/SCADA devices can be found too easily, change the default TCP/UDP port numbers of such devices and change the banners that identify the devices. This ensures that no unnecessary information about the device (such as product version and available modules) is revealed. Although this recommendation does *not* prevent discoverability of a device, it does make it harder.

- Use techniques to restrict network traffic on ports and protocols associated with ICS/SCADA services. Examples of such techniques are rate-limiting and the whitelisting of legitimate users. Restriction not only provides protection against potential hacking attempts, but also against Denial-of-Service (DoS) and/or brute force attacks.

- Harden the device configuration by disabling functionalities and services that are not used by the managers and operators. This process also includes removing unnecessary usernames or logins, changing default passwords and uninstalling unnecessary software and hardware modules. The goal is to reduce the potential attack surface by exposing only the necessary services.

- Maintain an up-to-date list of software and hardware that is running in your infrastructure. In this way it becomes easy to identify if newly discovered vulnerabilities may become a threat to your system.

- Monitor the manufacturer vulnerabilities. The manufacturers often directly contact their customers when a patch is available for their devices. However, subscribing to some known vulnerability databases, such as ICS-CERT and NVD is also recommended.

- Keep other systems that interact with the ICS/SCADA devices secure and ensure that they run the latest software version. Some attacks exploit weaknesses in adjacent systems, in order to bypass the imposed access restrictions.

- Monitor and assess the online discoverability and vulnerability of your ICS/SCADA devices. This report only provides a snapshot of the situation in 2018. Therefore we suggest organisations to follow the methodology described in this report and periodically check if (parts of) their infrastructure are found to be discoverability and even vulnerable. Organisations concerned about their security should consider the regular use of professional "security red-teams" that try to explore the vulnerabilities of devices within an ICS/SCADA infrastructure.

- Set-up a measurement and logging infrastructure, to detect possible scanning and attacking attempts in a stage as early as possible. Examples include Intrusion Detection Systems (IDSs) and flow-measurement systems.

- Ensure that the default passwords of ICS/SCADA devices are changed, since default passwords can be easily found on the Internet.

- In addition to their SCADA protocols, ICS/SCADA devices may have built-in web services for configuration and management purposes. Be aware of such services, and take appropriate actions to protect such services.

- DNS logs can be used to detect potentially unauthorised access. In the Netherlands we might consider whether organisations like SIDN, who maintains the DNS within the Netherlands, should play a role in such detection.

Monitoring if devices are discoverable and vulnerable is an important first step in protecting the ICS/SCADA infrastructure from unwanted access. However, it is important to stay aware of zero-day vulnerabilities/attacks, which can not been foreseen. This means that even when running the latest software/firmware versions, some devices will be susceptible to attacks.

Finally we recommended that a discussion gets started whether a *dedicated Trusted and Resilient network for the critical infrastructures* should be established. Such dedicated network could somehow resemble the existing 112 network and current ISPs could work together to establish such network. For the motivation behind this recommendation, see the discussion section of Chapter 6.

# 6
Chapter

**Highlights of this chapter:**

This report discusses the security of ICS/SCADA systems in the Netherlands and tackled three main questions:

- *How many ICS/SCADA devices located in the Netherlands can be easily found* by potential attackers?,
- *How many of these devices are vulnerable* to cyber attacks?,
- *What measures* should be taken to prevent these devices from being hacked?

The approach taken by this study was to start with a literature study to determine which well known ICS/SCADA protocols exist and what are their related ports numbers. The result of this literature study is a list of 39 protocols, which can be used as input to the Shodan search engine. To avoid false positives, two other lists were developed with features that would verify whether a device is indeed an ICS/SCADA device (positive), or some other kind of device (negative).

The main conclusions are that:

- tools like *Shodan* (see Chapter 2) make it extremely easy for potential attackers to find ICS/SCADA devices,
- *almost one thousand (989) ICS/SCADA devices in the Netherlands are exposed* on the Internet (see Chapter 3),
- around *sixty of these devices have multiple vulnerabilities* with a high severity level (see Chapter 4) and
- several well-known and relatively easy to deploy *measures* exist that help to improve the security of these ICS/SCADA devices (see Chapter 5).

These findings are worrying, particularly since the numbers provided in this study must be seen as lower bounds. Professional hackers, such as those working for criminal organisations and nation states, are certainly able to discover and hack more devices..

We therefore believe it is time to reconsider how critical infrastructures are connected to, and therefore exposed on the Internet. A discussion is needed whether it is time to establish a *dedicated Trusted and Resilient network for the critical infrastructures.*

*<This page was intentionally left blank.>*

## *6.1* Conclusions per chapter

The aim of this section is, for each of the previous chapters, to highlight the chapter's goals, summarise the methodologies, discuss the findings and draw conclusions.

- The goals of Chapter 1 (Introduction) were: 1) to introduce the motivation for the research in this report; 2) to provide a brief description of the methodologies used; and 3) to define the scope of the research.
    - The motivation underlying this research is the observation that ICS/SCADA devices have been inadvertently exposed on the public Internet without proper security measures, potentially causing catastrophic incidents.
    - Our approach involved several steps, including the collection of IP addresses of devices in the Netherlands, classifying these devices as ICS/SCADA devices or not, and identifying the known vulnerabilities of the classified devices. For these steps datasets were used from Shodan, ICS-CERT, and NVD. As part of this study a list was created of well known ICS/SCADA protocols and their related TCP/IP port numbers. For validation purposes two other lists were developed, one that would identify a device as ICS/SCADA device (positive), and another to identify a device as *non* ICS/SCADA device (negative).
    - This study could be extended by performing port scans ourselves, include the analysis of IPv6 addresses, investigate port numbers different from the default ICS/SCADA protocols, investigate ICS/SCADA protocols not yet contained on the list that we created, or use information from databases different from those provided by Shodan, ICS-CERT, and NVD. The outcome of such extended study would likely be a higher number of vulnerable devices. The numbers presented in this report should therefore be seen as lower bounds; the actual numbers may be higher.
- The goals of Chapter 2 (ISC/SCADA Device Discoverability), were: 1) to compose a comprehensive list of ICS/SCADA protocols and port numbers and 2) to describe a comprehensive and effective way to discover ICS/SCADA devices on the Internet.
    - Based on an extensive literature study, we composed a list of the 39 best known ICS/SCADA protocols, including their TCP/UDP port numbers (Table 2.1).
    - One possible approach to discover ICS/SCADA devices is to perform a port scan. However, scanning and fingerprinting the more than 4 billion IPv4 devices on the Internet brings technical, ethical, and legal issues (see Chapter 2 for details). Therefore we decided to use information obtained from an existing project that performs regular Internet-wide scans.
    - After comparing the best-known scanner projects, we concluded that Shodan provided the most comprehensive list of devices. We therefore decided in the remainder of this study to rely on Shodan for detecting devices. It should be noted, however, that relying on a single project possibly limits the number of results. The numbers provided by this report should therefore be considered as minimum numbers.
    - We observed that, to classify ICS/SCADA devices, it is necessary to have (1) the port number and (2) meta-data retrieved from the ICS/SCADA device. This meta-data is, in general, a configurable "welcome" message that the device returns when it is connected to. It usually contains system information, such as the operating system (OS), software/firmware versions, and Web services running on specific port numbers. Note that ICS/SCADA devices that did not respond with any meta-data, have not been classified as ICS/SCADA devices by this study.
- The goals of Chapter 3 (Exposed ICS/SCADA Devices in the Netherlands) were: 1) to explain the methodology used for classifying ICS/SCADA devices, and 2) to discuss the characteristics of the exposed ICS/SCADA devices.
    - Our methodology to classify ICS/SCADA devices relies on (1) the port number and (2) meta-data obtained from all devices (IP addresses) geolocated in the Netherlands. In Chapter 2 we described how we found these IP addresses. Connections to these IP addresses were created, and the returned meta-data was collected and analysed. In this analysis we compared the meta-data to

positive or negative keywords (Appendix B). These keywords are in fact the signatures of ICS/-SCADA devices (the list with positive features), or of devices that are known *not* to be ICS/SCADA devices (the list with negative features). Devices that did not return a signature, or returned an unknown signature, were not considered to be ICS/SCADA devices. For example, an IP address with port number 502 must have the word 'Modbus/TCP' in the meta-data, otherwise this IP address cannot be classified as an ICS/SCADA device. A limitation of our approach is that we cover only default ICS/SCADA ports, and that we can not analyse devices without meta-data information.

- In the Netherlands we found 3,09 million devices that are connected to the Internet. Almost one thousand (989) of these devices could be classified as ICS/SCADA devices. This number is substantial, considering that **anyone** connected to the Internet is able to access those devices. In Chapter 5 we provided some recommendations on how to minimise and protect ICS/SCADA devices.

- We observed that more than five hundred ICS/SCADA devices (557 devices), which represents 55.31% of all ICS/SCADA devices in the Netherlands, are related to the Tridium manufacturer. The next most popular manufacturer, Omron, accounts for five times fewer devices than Tridium (112 devices). One of the main explanations for the large number of Tridium devices relates to the nature of these devices. Tridium devices are generic and can be used in any sector. Additionally, most of the Tridium devices have the ability to interoperate with devices and protocols from other manufacturers.

- When investigating a device (that is, the exact service running on the device), again, a large majority of the ICS/SCADA systems were from Tridium. While looking at the products from other manufacturers, we noticed that many of these are used to connect legacy ICS/SCADA equipment to the Internet. Alarmingly, we observe that these devices do not have built-in security. We therefore advise managers and operators of ICS/SCADA systems to replace legacy equipment with more secure equipment.

- Finally, we investigated both physical and cyber locations of the ICS/SCADA devices. We were only able to identify the Internet Service Providers that host the discovered devices. This means that organisations running ICS/SCADA devices neither run their devices in their own Autonomous System nor have a dedicated infrastructure to forward their traffic. On the one hand this means that the organisations are hidden behind ISPs, however, on the other hand, the ISPs become critical points of failure for the organisations running the ICS/SCADA systems. For example, a Distributed Denial of Service attack on an ISP may also prevent the organisations using ICS/SCADA devices from managing their infrastructure.

- The goals of Chapter 4 (ICS/SCADA Devices Vulnerabilities in the Netherlands) were: 1) to explain a methodology to classify if ICS/SCADA devices are vulnerable or not and 2) to apply this methodology to find vulnerable ICS/SCADA devices in the Netherlands.

  - The approach to classify whether ICS/SCADA devices are vulnerable or not uses three pieces of meta-data collected from the ICS/SCADA devices: manufacturer, service, and service version. These pieces of meta-data are compared to two publicly available list of vulnerabilities: ICS-CERT and NVD. Although this approach is relatively straightforward and provides useful results, the downside is that vulnerabilities not included on one of these two lists can not be detected.

  - In addition to the classification of vulnerabilities, we propose a methodology for assessing the severity of vulnerabilities. For this assessment we rely on a method proposed by NIST, which categorises ranges of CVSS scores into three severity levels: low, medium, and high. Therefore, for each vulnerability that we identify, we also retrieve the severity level from ICS-CERT and NVD.

  - We observed that of the 989 ICS/SCADA devices in the Netherlands, only 6% (63) devices have one or more vulnerabilities. Although this number is relatively low, each of these devices can be easily exploited by hackers with unforeseeable consequences.

  - We also found that the majority of vulnerable ICS/SCADA devices have *multiple* vulnerabilities. For example, that are 8 devices that have 16 vulnerabilities each. There are several plausible explanations for this finding. For example, it could be that those vulnerable devices were not updated due to a lack of security awareness. Also it could be that certain companies prefer to manage the risk and avoid the possible instability caused by a software update. Yet another possibility is that the vulnerable devices are no longer part of a production infrastructure, and have been 'forgotten'.

  - While investigating the detailed list of vulnerabilities of ICS/SCADA devices in the Netherlands, we identified 37 vulnerabilities. Most of these vulnerabilities can easily be solved by updating the

- software of the device. As explained above, there may be multiple explanations why such updates didn't happen.
- While investigating the severity level of vulnerabilities, we observed that 83% were classified with high severity, 9% as medium and 8% as low severity. Note that most of the ICS/SCADA devices classified as vulnerable had more than one vulnerability. We notice that most of the devices have at least one vulnerability with a high level of severity. This finding emphasises that all 63 devices are vulnerable and should be patched.
- While investigating the vendors and product related to the ICS/SCADA devices , we observed that the vulnerable devices come from only five vendors: Omron, Siemens, Rockwell, Schneider, and Tridium. This finding does not mean that these five manufacturers are 'insecure', but that organisations that use ICS/SCADA devices from these vendors seem to be reluctant to deploy patches that would improve security.
- Finally, while investigating in which organisations and sectors the vulnerable ICS/SCADA devices are deployed, we were only able to obtain the name of the ISP via which these devices are connected to the Internet (see also § 3.2.4). As may be expected, we observed that the most important ISPs in the Netherlands (KPN, ZIGGO, VODAFONE and LIBERTY GLOBAL) also host most of the vulnerable devices.

- The goals of Chapter 5 (measures) were to propose measures and recommendations to improve the security of ICS/SCADA devices in the Netherlands .

  - The measures to improve the security of ICS/SCADA devices are basically well-known, easy to implement and in-line with the 'Checklist security of ICS/SCADA systems' issued by the National Cyber Security Centre (NCSC).
  - To prevent the discovery of ICS/SCADA devices using search engines such as Shodan, ICS/SCADA devices should no longer be directly accessible from the open Internet, but hidden behind firewalls, Virtual Private Networks (VPNs) and/or Virtual Local Area Networks (VLANs).
  - Techniques should be applied that restrict network traffic on ports and protocols associated with ICS/SCADA services. Examples of such techniques are rate-limiting and the whitelisting of legitimate users. Restriction not only provides protection against potential hacking attempts, but also against Denial-of-Service (DoS) and/or brute force attacks.
  - Software updates should be installed in a timely manner. When it is not feasible to update the software, make sure the device can not be accessed via the Internet.

## *6.2* Discussion

At the end of this study the question raises whether the findings are good or bad. In other words, are critical infrastructures in the Netherlands in danger or not? Is a scenario like the December 2015 Ukraine power grid attack, which left more than two hundred thousand people without electricity for multiple hours, possible in the Netherlands or not? The answer to this question is mixed.

One possible answer is that in the Netherlands almost one thousand (989) ICS/SCADA devices can be found quite easily with tools like Shodan, and that around sixty of them show multiple vulnerabilities with a high severity level. If an attacker is able to take over control of just a single device, the consequences *may in theory be* the failure of a complete critical infrastructure, such as a lock gate or power plant. Even worse, the methods and tools used in this study are relatively simple and can already be used by script kiddies and amateur hackers. Therefore the numbers provided in this study must be seen as lower bounds. Professional hackers, such as those working for criminal organisations and nation states, are certainly able to discover and hack more devices. The reasons that the real numbers will be higher are:

- This study was limited to the use of Shodan, and did not check the outcome of other search engines, such as Censys, or other scanning projects. An interesting follow-up study would therefore be to compare the outcome of other search engines and scanning projects to the results obtained by this study. The expected outcome is that more (vulnerable) devices will be found.

- This study was limited to well-known TCP/UDP ports that are used by ICS/SCADA protocols. Professional hackers could build dedicated ICS/SCADA scanners, that first determine which IP address range belongs to a certain critical infrastructure, and subsequently scan *all* TCP/UDP ports within that range, instead of just the well-known ports. An interesting follow-up study would therefore be to investigate if professional hackers already use such dedicated scanners. For such study one would need to create, within the IP address space of critical infrastructure providers, "telescopes" and "honeypots", which record and analyse scanning attempts.

- This study was limited to the traditional IPv4 address space, which has 32-bit addresses, allowing $2^{32}$ (4,294,967,296) systems to be addressed. With standard scanning techniques, such as used in this study, it is well possible to scan (brute force) the entire IPv4 address space.
  ICS/SCADA devices may also be connected via IPv6, however. The IPv6 address space supports 128-bit addresses and therefore allows $2^{128}$ (340,282,366,920,938,463,463,374,607,431,768,211,456) systems to be addressed. This number is too big to scan brute-force (that was also the reason why this study did not take the IPv6 space into account). However, by using sophisticated scanning techniques that focus on small parts of the IPv6 address space, professional attackers may still be able to find ICS/SCADA devices within the IPv6 address space [49].

- Using one or more of the suggestions provided above, professional hackers would certainly be able to discover more ICS/SCADA devices than the 989 reported by this study. But not only would they be able to find more *connected devices*, by using zero-day exploits they will also be able to find more *vulnerable devices*, since this study restricted itself to *known vulnerabilities* only.

Another possible answer is that this study did not investigate for *which purpose* the devices that have been found are being used, nor the *actual impact* that a hack of one of such devices would have. Therefore it is, in principle, well possible that all Dutch critical infrastructures are completely secure.

Further study is thus needed to better understand for which purpose the (vulnerable) ICS/SCADA devices found in this study are being used. In fact it is well possible that, at least a substantial number of the (vulnerable) devices found in this study, are not, or no longer connected to critical infrastructures, but instead used for non-critical applications, or as stand-alone devices for testing purposes. In fact, we know that at least some of the devices found in this study are used for research purposes. Although unlikely, it is even possible that the (vulnerable) devices found in this study are part of a honeypot to attract, identify and isolate hackers.

An interesting question therefore is to discover which organisations are responsible for operating the devices found in this study. This study revealed only the IP addresses of devices, and not the organisation behind these devices. To map the addresses of these devices to organisations, help from ISPs that connect these devices to the Internet would be desirable. However, for privacy (and security) reasons, ISPs are not allowed to provide these mappings to researchers. Therefore other approaches are needed:

- Instead of investigating whether the IP addresses found in this study belong to a critical infrastructure, the providers of critical infrastructures could obtain our list of IP addresses and check whether some of these addresses belong to them. The critical infrastructure providers can than check themselves if their devices are on the list, and determine the impact that a potential hack could have, and take appropriate mitigation actions. The various providers of critical infrastructures are organised within so-called Information Sharing and Analysis Centres (ISAC's) of the Dutch National Cyber Security Centre (NCSC), and our main recommendation is to share the findings of this study with them.

- Another approach would be to investigate the Domain Name System (DNS), in an attempt to map IP addresses to domain names and via these names obtain information regarding the organisations behind. However, a small feasibility study that was already performed as part of this research did not immediately reveal hopeful results. Still some results may be expected from a more extensive study. To determine whether these domains are indeed being queried, collaboration with SIDN could also be considered.

Additional study is also needed to assess the *actual impact* that a hack of the ICS/SCADA devices discovered in this study would have. For obvious reasons such hacking attempts can not and should not be performed without the explicit consent of the organisations operating these devices. Commercial pen-testing companies exist, however, that can perform such assessments on request. Although outside the scope of this study, it is assumed that already many of the critical infrastructure operators use the services of such companies, or are able to perform such assessments themselves. In addition, it is certainly possible that the organisations operating these devices are well aware of the vulnerabilities within their devices, but decided that the real risks are too low to justify immediate action.

Our final recommendation is that a discussion gets started whether a *dedicated Trusted and Resilient network for the critical infrastructures* should be created. As discussed in Chapter 3, critical infrastructure organisations seem to rely for routing and protection on general ISP services. Failure of such services, for example as a result of a massive Distributed Denial of Service (DDoS) attack, could have severe consequences for the operation and management of critical infrastructures. In recent years we have witnessed a sharp increase in the number and power of DDoS attacks. Even though ISPs, together with others, make good progress in the protection against DDoS attacks (Dutch Continuity Board, Anti-DDoS Working Group), we should still prepare for potentially far stronger attacks initiated by criminal organizations or even nation states. By creating a *dedicated Trusted and Resilient network for the critical infrastructures*, protection of such infrastructures could become more effective. Such protection would not only be against DDoS attacks, but also against the discoverability and the exploitation of device vulnerabilities. Current ISPs could work together to establish such network, which would somehow resemble the existing 112 network.

*<This page was intentionally left blank.>*

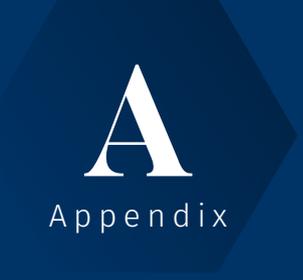

# ICS/SCADA Protocols and Search Engines

This table presents the most common ICS/SCADA Protocols and their coverage by search engines. In the table the protocols that were used in our queries to the search engines have been highlighted. It is important to note that these protocols were used as an initial filter to search for devices. However, as described in our methodology, for each device that matched our search criteria, we enriched the information with that of other ports belonging to the same IP address. In that way, we were able to map devices that are running other protocols. For example, the protocol `SAIA S-BUS` is not supported by Shodan. However, when the information of devices was enriched in conjunction with the positive keywords, we managed to include them.

Table A.1: Well known ICS/SCADA protocols and how these are covered by scanning projects.

|    | Protocol | Shodan | Censys |
|----|----------|--------|--------|
| 1  | ANSI C12.22 | no | no |
| 2  | BACNet | yes | yes |
| 3  | Beckhoff-ADS communication | no | no |
| 4  | CANopen | no | no |
| 5  | CodeSys | yes | no |
| 6  | Crimson 3 | yes | no |
| 7  | DNP3 | yes | yes |
| 8  | Danfoss ECL apex | no | no |
| 9  | EtherCAT | no | no |
| 10 | EtherNet/IP | yes | no |
| 11 | FATEK FB Series | no | no |
| 12 | GE-SRTP | yes | no |
| 13 | HART-IP | yes | no |
| 14 | HITACHI EHV Series | no | no |
| 15 | ICCP | no | no |
| 16 | IEC 60870-5-104 | yes | no |
| 17 | IEC 61850 / MMS | yes | no |
| 18 | KEYENCE KV-5000 | no | no |
| 19 | KOYO Ethernet | no | no |
| 20 | LS Fenet | no | no |
| 21 | MELSEC Q | yes | no |
| 22 | Modbus/TCP | yes | yes |
| 23 | Moxa | yes | no |
| 24 | Niagara Tridium Fox | yes | no |
| 25 | OMRON FINS | yes | no |
| 26 | OPC | no | no |
| 27 | PCWorx | yes | no |
| 28 | Panasonic FP (Ethernet) | no | no |
| 29 | Panasonic FP2 (Ethernet) | no | no |
| 30 | ProConOS | yes | no |
| 31 | Quick Panel GE | no | no |
| 32 | SAIA S-BUS (Ethernet) | no | no |
| 33 | Schleicher XCX 300 | no | no |
| 34 | Siemens S7 | yes | yes |
| 35 | Simatic | no | no |
| 36 | Unitronics Socket1 | no | no |
| 37 | YASKAWA MP Series Ethernet | no | no |
| 38 | YASKAWA MP2300Siec | no | no |
| 39 | Yokogawa FA-M3 (Ethernet) | no | no |

This table describes the most popular protocols and their coverage. Moreover, Shodan uses a set of signatures to determine the device's manufacturer. These signatures aim to enrich the information provided by classifying the device protocol and the respective manufacturer. For example, Shodan can identify Modbus devices from Siemens, Crestron, Schneider, and others.

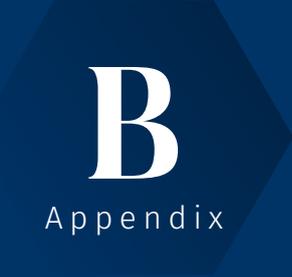

# Appendix B

# Features for ICS/SCADA Device Classification

# Positive Features for Classifying ICS/SCADA

The table below shows the keywords that have been used to validate if a device is an ICS/SCADA device. We compare the keywords of this list to the meta-data received from the device. This meta-data is generally some kind of 'welcome' message that the device returns once a connection is made. If the meta-data contains keywords from the table below, we can be sure the devices is an ICS/SCADA device (true positive).

Table B.1: Positive features related to ICS/SCADA devices.

|    | Positive Features |
|----|-------------------|
| 1  | ABB               |
| 2  | RVT               |
| 3  | PLC               |
| 4  | LINX              |
| 5  | ACTL              |
| 6  | Moxa              |
| 7  | Wago              |
| 8  | CJ1M              |
| 9  | Ewon              |
| 10 | CJ2M              |
| 11 | CP2E              |
| 12 | CJ2H              |
| 13 | Omron             |
| 14 | E-DDC             |
| 15 | Sauter            |
| 16 | Loytec            |
| 17 | Gaspot            |
| 18 | PXG3.L            |
| 19 | Sirius            |
| 20 | OJ-Air2           |
| 21 | Niagara           |
| 22 | Siemens           |
| 23 | Phoenix           |
| 24 | L26CPU            |
| 25 | Creston           |
| 26 | Tridium           |
| 27 | E-MIO C           |
| 28 | E-MIO H           |
| 29 | L06CPU            |
| 30 | L02CPU            |
| 31 | X-MIO1.0          |
| 32 | Crestron          |
| 33 | E-DDC6.3          |
| 34 | Rockwell          |
| 35 | Q02UCPU           |
| 36 | Q01UCPU           |
| 37 | SIMOTION          |
| 38 | Q00UCPU           |
| 39 | 3s-smart          |
| 40 | 1763-BA           |
| 41 | L02SCPU           |
| 42 | MoxaHttp          |
| 43 | MELSEC-Q          |
| 44 | MELSEC-L          |
| 45 | AB Regin          |
| 46 | 1766-MM1          |
| 47 | 1768-CNB          |
| 48 | 1769-PA4          |
| 49 | 1769-PA2          |
| 50 | L26CPU-P          |

| # | Model |
|---|---|
| 51 | 1768-PB3 |
| 52 | Lantronix |
| 53 | 1763-MM1 |
| 54 | L02CPU-P |
| 55 | Q00UJCPU |
| 56 | L06CPU-P |
| 57 | X-RUS 3.0 |
| 58 | Omron PLC |
| 59 | Q03UDCPU |
| 60 | 1769-PB2 |
| 61 | X-SRCO2 T |
| 62 | Schneider |
| 63 | 1756-L72 |
| 64 | S-943460 |
| 65 | 1756-L63 |
| 66 | Solar-Log |
| 67 | 1756-L73 |
| 68 | 1756-L61 |
| 69 | 1756-L64 |
| 70 | 1756-L74 |
| 71 | 1756-L62 |
| 72 | PCD2.M5..0 |
| 73 | PCD1.M2XXX |
| 74 | X-ERW3 ANT |
| 75 | PCD3.M..60 |
| 76 | PCD3.M..x0 |
| 77 | PCD1.M2..0 |
| 78 | PCD1.M0..0 |
| 79 | 1756-L83E |
| 80 | 1756-L72K |
| 81 | PCD3.Mxxx7 |
| 82 | Mitsubishi |
| 83 | 1756-L82E |
| 84 | 1756-L72S |
| 85 | L02SCPU-P |
| 86 | 1756-L7SP |
| 87 | TM221CE16R |
| 88 | 1756-L74K |
| 89 | TM221CE40T |
| 90 | L26CPU-BT |
| 91 | TM221CE40R |
| 92 | Q20UDHCPU |
| 93 | PCD7.D4..D |
| 94 | PCD7.D4..V |
| 95 | 1756-L84E |
| 96 | Q13UDVCPU |
| 97 | Q13UDHCPU |
| 98 | Q10UDHCPU |
| 99 | 1756-L61S |
| 100 | Q26UDHCPU |
| 101 | 1763-NC01 |
| 102 | Q06UDVCPU |
| 103 | Q06UDHCPU |
| 104 | Q04UDVCPU |
| 105 | Q04UDHCPU |
| 106 | 1756-L73K |
| 107 | 1768-ENBT |
| 108 | Q26UDVCPU |
| 109 | 1756-L73S |
| 110 | 1756-L62S |
| 111 | Q03UDVCPU |
| 112 | Q03UDECPU |
| 113 | 1756-L63S |
| 114 | 1768-EWEB |
| 115 | 1756-L8SP |
| 116 | 1756-L85E |
| 117 | CJ1H_CPU67H |
| 118 | 1756-ENBTK |
| 119 | CJ1G_CPU45H |
| 120 | 1756-L63XT |
| 121 | CS1G_CPU44H |
| 122 | 1756-L71SK |
| 123 | CS1G_CPU45H |
| 124 | GuardLogix |
| 125 | BC-FM-BSK24 |
| 126 | CS1H_CPU66H |
| 127 | 1756-L72SK |
| 128 | Q20UDEHCPU |
| 129 | 1756-L73XT |
| 130 | Q26UDEHCPU |
| 131 | Q13UDEHCPU |
| 132 | Q10UDEHCPU |
| 133 | Q06UDEHCPU |
| 134 | Q50UDEHCPU |
| 135 | RUS 2.1 ANT |
| 136 | 1756-RM2XT |
| 137 | 1756-L8SPK |
| 138 | 1756-L73SK |
| 139 | 1756-L84EK |
| 140 | 1769-L30ER |
| 141 | 1756-L84ES |
| 142 | 1756-L83ES |
| 143 | 1756-L83EK |
| 144 | Niagara Fox |
| 145 | 1756-L82ES |
| 146 | 1756-L82EK |
| 147 | 1756-L81EK |
| 148 | 1756-L7SPK |
| 149 | LIBIEC61850 |
| 150 | L26CPU-PBT |
| 151 | 1769-L33ER |
| 152 | Q04UDEHCPU |
| 153 | PCD1.M2110R1 |
| 154 | Satel-iberia |
| 155 | EY-RC500F001 |
| 156 | EY-AS525F001 |
| 157 | Q100UDEHCPU |
| 158 | 1756-ENBT/A |
| 159 | 1769-L38ERM |
| 160 | 1763-L16AWA |
| 161 | 1763-L16DWD |
| 162 | 1762-L40BXB |
| 163 | 1766-L32AWA |
| 164 | 1769-L37ERMS |
| 165 | 1766-L32BWA |
| 166 | 1762-L40BWA |
| 167 | 1769-L37ERM |
| 168 | 1762-L40AWA |
| 169 | 1766-L32BXB |
| 170 | 1762-L24BXB |
| 171 | 1763-L16BBB |
| 172 | BMX NOE 0100 |
| 173 | 1762-L24BWA |
| 174 | BC-FM-BSK230 |
| 175 | 1763-L16BWA |
| 176 | 1756-L84ESK |
| 177 | 1756-L83ESK |
| 178 | 1756-L82ESK |
| 179 | 1769-L33ERM |
| 180 | BMX P34 2020 |
| 181 | 1756-L81ESK |
| 182 | 1769-L36ERM |
| 183 | CompactLogix |
| 184 | 1756-L73SXT |
| 185 | 1769-L30ERM |
| 186 | 1769-L33ERMK |
| 187 | 1769-L33ERMO |
| 188 | 1766-L32AWAA |
| 189 | 1766-L32BXBA |
| 190 | 1769-L37ERMOS |
| 191 | 1769-L37ERMO |
| 192 | 1769-L36ERMO |
| 193 | ControlLogix |
| 194 | 1766-L32BWAA |
| 195 | 1769-L36ERMS |
| 196 | 1769-L33ERMS |
| 197 | 1769-L37ERMK |
| 198 | PCD7.D4..WTPF |
| 199 | 1769-L37ERMS |
| 200 | 1756-L72EROM |
| 201 | CompactLogix |
| 202 | 1756-L73EROM |
| 203 | BMX P34 20302 |
| 204 | SE-Elektronic |
| 205 | 1762-L40BXBR |
| 206 | 1762-L24BWAR |
| 207 | 1762-L24BXBR |
| 208 | 1769-L30ERMS |
| 209 | 1762-L40AWAR |
| 210 | 1769-L38ERMS |
| 211 | 1769-L38ERMOS |
| 212 | 1769-L38ERMO |
| 213 | 1769-L38ERMK |
| 214 | 1769-L30ERMK |
| 215 | 1762-L40BWAR |
| 216 | Allen-Bradley |
| 217 | 1769- L30ERMS |
| 218 | Simatic S7-300 |
| 219 | 1756-L73EROMS |
| 220 | 1756-L72EROMS |
| 221 | SCALANCE M-800 |
| 222 | CP1L-EL20DT1-D |
| 223 | 1769-L33ERMOS |
| 224 | 1769-L38ERMSK |
| 225 | 1769-L37ERMSK |
| 226 | ILC 131 ETH/XC |
| 227 | OJ Electronics |
| 228 | CP1L-EM30DT1-D |
| 229 | 1769-L36ERMOS |
| 230 | 1769-L33ERMSK |
| 231 | GuardLogix 5370 |
| 232 | ILC 171 ETH 2TX |
| 233 | 1769-L30ER-NSE |
| 234 | MicroLogix 1400 |
| 235 | Mitsubishi Q PLC |
| 236 | MicroLogix 1100 |
| 237 | MicroLogix 1200 |
| 238 | MicroLogix 1400 |
| 239 | ILC 151 GSM/GPRS |
| 240 | ILC 150 GSM/GPRS |
| 241 | Wago Corporation |
| 242 | ArmorGuardLogix |
| 243 | GuardLogix 5370 |
| 244 | 1769-L16ER-BB1B |
| 245 | 1769-L18ER-BB1B |
| 246 | 1769-L19ER-BB1B |
| 247 | 1769-L24ER-QB1B |
| 248 | 1756-L651756-L71 |
| 249 | CompactLogix 5370 |
| 250 | Red Lion Controls |
| 251 | Omron Corporation |
| 252 | 1769-L18ERM-BB1B |
| 253 | NiagaraAX Station |
| 254 | 1756-LSP1756-L71S |
| 255 | 1756-L751756-L71K |
| 256 | ArmorControlLogix |
| 257 | CompactLogix 5370 |
| 258 | 1769-L27ERMQBFC1B |
| 259 | 1769-PB41756-ENBT |
| 260 | 1769-L24ER-QBFC1B |
| 261 | 1768-CNBR1768-PA3 |
| 262 | 1769-L24ER-QBFC1BK |
| 263 | 1756-L75K1756-L81E |
| 264 | PCD3.M5547 1.08.33 |
| 265 | Armor CompactLogix |
| 266 | GNU Lib LIBIEC61850 |
| 267 | Rockwell Automation |
| 268 | SIMATIC HMI Comfort |
| 269 | 1756-L85EK1756-RM2 |
| 270 | Tridium Niagara httpd |
| 271 | Saia Burgess Controls |
| 272 | General Electric SRTP |
| 273 | 1756-L7SPXT1756-L81ES |
| 274 | MicroLogix 1400 FRN 21 |
| 275 | MicroLogix 1400 FRN 21 |
| 276 | Siemens HiPath 3000 telnetd |
| 277 | 3S-Smart Software Solutions |
| 278 | Solare Datensysteme GmbH V1.00 |
| 279 | PXC64-U \+ PXA30-W0 / HW=V2.02 |
| 280 | Satel ETHM-1 alarm control unit |
| 281 | PXC50-E.D \+ PXA40-W0 / HW=V1.00 |
| 282 | CP3 Console 3-Series Control System |
| 283 | CP2 Console 2-Series Control System |
| 284 | Building Operation Automation Server |
| 285 | MicroLogix 1400 FRN 21 1766-L32BWA B |
| 286 | MicroLogix 1400 FRN 21 1766-L32AWAA B |
| 287 | PXC100-E.D \+ PXA40-W0 \+ PXX-PBUS / HW=V3.00 |
| 288 | TAC Xenta 555 programmable logic controller httpd |
| 289 | TAC Xenta 511 programmable logic controller httpd |

# Negative Features for Classifying ICS/SCADA

The table below shows the keywords that have been used to validate if a device is *not* an ICS/SCADA device. We compare the keywords of this list to the meta-data received from the device. This meta-data is generally some kind of 'welcome' message that the device returns once a connection is made. If the meta-data contains keywords from the table below, we can be sure the devices is *not* an ICS/SCADA device (false positive).

Table B.2: Negative features related to ICS/SCADA devices.

|    | Negative Features |
|----|-------------------|
| 1  | RFB |
| 2  | .NET |
| 3  | LDAP |
| 4  | POP3 |
| 5  | IMAP |
| 6  | IRCd |
| 7  | TFTP |
| 8  | SMTP |
| 9  | smtp |
| 10 | FTP |
| 11 | FTPd |
| 12 | RTSP |
| 13 | http |
| 14 | HTTP |
| 15 | Squid |
| 16 | nginx |
| 17 | mysql |
| 18 | Hydra |
| 19 | Monero |
| 20 | EdgeOS |
| 21 | Ubuntu |
| 22 | Apache |
| 23 | RSYNCD |
| 24 | Conpot |
| 25 | Debian |
| 26 | 220 FTP |
| 27 | NETBios |
| 28 | OpenSSH |
| 29 | WinSSHD |
| 30 | FlexHub |
| 31 | SSH-2.0 |
| 32 | Dovecot |
| 33 | Postfix |
| 34 | NetBIOS |
| 35 | ProFTPD |
| 36 | PowerDNS |
| 37 | HTTP/1.0 |
| 38 | HTTP/1.1 |
| 39 | VerliHub |
| 40 | get lost |
| 41 | WAR-FTPD |
| 42 | HTML 2.0 |
| 43 | SSH-1.99 |
| 44 | Minecraft |
| 45 | FileZilla |
| 46 | VPN (IKE) |
| 47 | DHT Nodes |
| 48 | Pure-FTPd |
| 49 | FTP server |
| 50 | ICY 200 OK |
| 51 | SMB Status |
| 52 | neptunehub |
| 53 | BGP Message |
| 54 | FTP service |
| 55 | WD My Cloud |
| 56 | Rust Server |
| 57 | Conan Exiles |
| 58 | DarkRP Server |
| 59 | Initiator SPI |
| 60 | Counter-Strike |
| 61 | FTP(S) Service |
| 62 | Via: SIP/2.0/UDP |
| 63 | HttpOnly;expires |
| 64 | Deathmatch Server |
| 65 | MailEnable Service |
| 66 | DirectAdmin Daemon |
| 67 | Microsoft ESMTP MAIL |
| 68 | ESMTP SendPulse SMTP |
| 69 | Microsoft FTP Service |
| 70 | ARK: Survival Evolved |
| 71 | Universal DDE Connector |
| 72 | reJailBreak 0.1a Server |
| 73 | Unable to get printer printer |
| 74 | \\\x15\\\x03\\\x01\\\x00\\\x02\\\x02\\\ |
| 75 | No connection is available now. Try again later! |
| 76 | The firewall on this server is blocking your |
| 77 | Invalid status 71 # status 71 - Apache Thrift |
| 78 | Uw verbinding naar deze server is geblokkeerd |
| 79 | Error code explanation: 400 = Bad request synt |

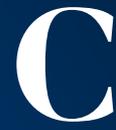

# ICS/SCADA Devices
# in the Netherlands

# ICS/SCADA Devices found in the Netherlands

This table describes all ICS/SCADA products found in the Netherlands. The first column describes the product's name, the second column the number of devices found, and the last column the percentage.

Table C.1: Products used in NL ICS/SCADA.

|    | Product | # Devices | Percentage |
|----|---------|-----------|------------|
| 1  | Niagara Fox | 545 | 45.45% |
| 2  | Tridium Niagara httpd | 195 | 16.26% |
| 3  | Moxa Nport | 67 | 5.59% |
| 4  | ILC 151 GSM/GPRS | 56 | 4.67% |
| 5  | Omron PLC | 56 | 4.67% |
| 6  | 3S-Smart Software Solutions | 36 | 3.00% |
| 7  | CJ2M | 23 | 1.92% |
| 8  | Simatic S7-300 | 18 | 1.50% |
| 9  | ILC 150 GSM/GPRS | 11 | 0.92% |
| 10 | CJ1M | 11 | 0.92% |
| 11 | MoxaHttp | 7 | 0.58% |
| 12 | MicroLogix 1400 FRN 21 1766-L32BWA B | 7 | 0.58% |
| 13 | E-DDC | 6 | 0.50% |
| 14 | EY-AS525F001 | 5 | 0.42% |
| 15 | ACTL | 5 | 0.42% |
| 16 | CS1G_CPU44H | 4 | 0.33% |
| 17 | CS1G_CPU45H | 4 | 0.33% |
| 18 | Lantronix | 4 | 0.33% |
| 19 | CJ2H | 3 | 0.25% |
| 20 | CJ1G_CPU45H | 3 | 0.25% |
| 21 | Solare Datensysteme GmbH V1.00 | 3 | 0.25% |
| 22 | SCALANCE M-800 | 3 | 0.25% |
| 23 | TAC Xenta 511 programmable logic controller httpd | 2 | 0.17% |
| 24 | NiagaraAX Station | 2 | 0.17% |
| 25 | BMX P34 20302 | 2 | 0.17% |
| 26 | SIMATIC HMI Comfort | 2 | 0.17% |
| 27 | Siemens HiPath 3000 telnetd | 2 | 0.17% |
| 28 | CP1L-EL20DT1-D | 2 | 0.17% |
| 29 | TM221CE40R | 2 | 0.17% |
| 30 | Omron Corporation | 2 | 0.17% |
| 31 | BMX NOE 0100 | 2 | 0.17% |
| 32 | Rockwell Automation/Allen-Bradley | 1 | 0.08% |
| 33 | CJ1H_CPU67H | 1 | 0.08% |
| 34 | SIMOTION | 1 | 0.08% |
| 35 | Satel ETHM-1 alarm control unit | 1 | 0.08% |
| 36 | PXG3.L | 1 | 0.08% |
| 37 | TAC Xenta 555 programmable logic controller httpd | 1 | 0.08% |
| 38 | TM221CE16R | 1 | 0.08% |
| 39 | TM221CE40T | 1 | 0.08% |
| 40 | Wago Corporation | 1 | 0.08% |
| 41 | RVT | 1 | 0.08% |
| 42 | Building Operation Automation Server | 1 | 0.08% |
| 43 | PCD1.M2XXX | 1 | 0.08% |
| 44 | BMX P34 2020 | 1 | 0.08% |
| 45 | CP1L-EM30DT1-D | 1 | 0.08% |
| 46 | CP3 Console 3-Series Control System | 1 | 0.08% |
| 47 | CS1H_CPU66H | 1 | 0.08% |
| 48 | CompactLogix | 1 | 0.08% |
| 49 | ControlLogix | 1 | 0.08% |
| 50 | EY-RC500F001 | 1 | 0.08% |
| 51 | General Electric SRTP | 1 | 0.08% |
| 52 | ILC 131 ETH/XC | 1 | 0.08% |
| 53 | ILC 171 ETH 2TX | 1 | 0.08% |
| 54 | LIBIEC61850 | 1 | 0.08% |
| 55 | LINX | 1 | 0.08% |
| 56 | MicroLogix 1400 FRN 21 1766-L32AWAA B | 1 | 0.08% |
| 57 | Mitsubishi Q PLC | 1 | 0.08% |
| 58 | OJ-Air2 | 1 | 0.08% |
| 59 | MELSEC-L | 1 | 0.08% |
| 60 | unknown | 81 | 6.76% |
| | TOTAL | 1.199 | 100.00% |

*<This page was intentionally left blank.>*

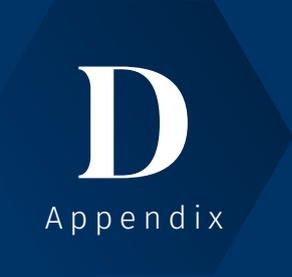

# ICS/SCADA Device manufacturers in the Netherlands

# Manufacturers of NL ICS/SCADA Devices

This table provides the vendors of the ICS/SCADA devices found in the Netherlands. The first column gives the name of the product manufacturer, the second column provides the number of devices found, and the last column the percentage.

Table D.1: Manufacturers of NL ICS/SCADA devices.

|    | Manufacturer          | # Devices | Percentage |
|----|-----------------------|-----------|------------|
| 1  | Tridium               | 557       | 55.31%     |
| 2  | Omron                 | 112       | 11.12%     |
| 3  | Phoenix               | 69        | 6.85%      |
| 4  | Moxa                  | 67        | 6.65%      |
| 5  | 3s-smart              | 36        | 3.57%      |
| 6  | Siemens               | 30        | 2.98%      |
| 7  | Schneider             | 13        | 1.29%      |
| 8  | Rockwell              | 11        | 1.09%      |
| 9  | Sauter                | 6         | 0.60%      |
| 10 | SE-Elektronic         | 6         | 0.60%      |
| 11 | Ewon                  | 5         | 0.50%      |
| 12 | Lantronix             | 4         | 0.40%      |
| 13 | Creston               | 3         | 0.30%      |
| 14 | Solar-Log             | 3         | 0.30%      |
| 15 | AB Regin              | 2         | 0.20%      |
| 16 | Mitsubishi            | 2         | 0.20%      |
| 17 | OJ Electronics        | 1         | 0.10%      |
| 18 | GE                    | 1         | 0.10%      |
| 19 | Saia Burgess Controls | 1         | 0.10%      |
| 20 | Satel-iberia          | 1         | 0.10%      |
| 21 | Loytec                | 1         | 0.10%      |
| 22 | GNU Lib LIBIEC61850   | 1         | 0.10%      |
| 23 | ABB                   | 1         | 0.10%      |
| 24 | Wago                  | 1         | 0.10%      |
| 25 | Unknown               | 73        | 7.25%      |
|    | TOTAL                 | 1.007     | 100.00%    |

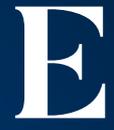

Autonomous Systems and ICS/SCADA Devices

# ASes Related to NL ICS/SCADA Devices

The table below shows via which Internet Service Providers (ISPs) the ICS/SCADA devices discovered in this study are connected to the Internet. The ISPs can be identified via their Autonomous System (AS) number. The first column shows the name associated with the AS (usually an ISP), the second column gives the actual AS number, the third column the number of ICS/SCADA devices that are reachable via this ISP, and the last column the percentage.

Table E.1: ASes related to NL ICS/SCADA devices.

|    | AS name             | AS number | count | percent |
|----|---------------------|-----------|-------|---------|
| 1  | KPN                 | AS1136    | 160   | 16.18%  |
| 2  | undefined           | AS9143    | 142   | 14.36%  |
| 3  | PT                  | AS8737    | 109   | 11.02%  |
| 4  | VFNL                | AS15480   | 80    | 8.09%   |
| 5  | ROUTIT              | AS28685   | 65    | 6.57%   |
| 6  | XS4ALL              | AS3265    | 64    | 6.47%   |
| 7  | LGI-UPC             | AS6830    | 56    | 5.66%   |
| 8  | SOLCON              | AS12414   | 30    | 3.03%   |
| 9  | REDHOSTING          | AS39647   | 24    | 2.43%   |
| 10 | SURFNET             | AS1103    | 20    | 2.02%   |
| 11 | ZEELANDNET          | AS15542   | 17    | 1.72%   |
| 12 | UNET                | AS29396   | 17    | 1.72%   |
| 13 | TMO                 | AS31615   | 13    | 1.31%   |
| 14 | TNF                 | AS33915   | 13    | 1.31%   |
| 15 | XENOSITE            | AS15426   | 12    | 1.21%   |
| 16 | PREVIDER            | AS20847   | 9     | 0.91%   |
| 17 | KABELFOON           | AS15435   | 9     | 0.91%   |
| 18 | VERSATEL            | AS13127   | 8     | 0.81%   |
| 19 | XXLNET              | AS34373   | 7     | 0.71%   |
| 20 | SPEEDXS             | AS30925   | 6     | 0.61%   |
| 21 | FIBERRING           | AS38930   | 6     | 0.61%   |
| 22 | TELECITY-LON        | AS15830   | 5     | 0.51%   |
| 23 | SERVERIUS           | AS50673   | 5     | 0.51%   |
| 24 | BBNED               | AS15670   | 5     | 0.51%   |
| 25 | SIGNET              | AS28878   | 5     | 0.51%   |
| 26 | undefined           | AS39309   | 5     | 0.51%   |
| 27 | NGNETWORKS          | AS57795   | 5     | 0.51%   |
| 28 | KPN-INTERNEDSERVICES| AS15879   | 4     | 0.40%   |
| 29 | NCBV-BACKBONE       | AS50554   | 4     | 0.40%   |
| 30 | EUROFIBER           | AS39686   | 3     | 0.30%   |
| 31 | ATOM86              | AS8455    | 3     | 0.30%   |
| 32 | QSP                 | AS12315   | 3     | 0.30%   |
| 33 | COLOCENTER          | AS58291   | 3     | 0.30%   |
| 34 | TMOBILE-THUIS       | AS50266   | 3     | 0.30%   |
| 35 | COMSAVE             | AS202120  | 3     | 0.30%   |
| 36 | INFOPACT            | AS21221   | 3     | 0.30%   |
| 37 | NEXTPERTISE         | AS41960   | 2     | 0.20%   |
| 38 | PLINQ               | AS35224   | 2     | 0.20%   |
| 39 | EQUINIXN            | AS47886   | 2     | 0.20%   |
| 40 | I3DNET              | AS49544   | 2     | 0.20%   |
| 41 | EURONET             | AS5390    | 2     | 0.20%   |
| 42 | BREEDBANDNEDERLAND  | AS5524    | 2     | 0.20%   |
| 43 | AS-TUE              | AS1161    | 2     | 0.20%   |
| 44 | DDF                 | AS35467   | 2     | 0.20%   |
| 45 | INTERCONNECT        | AS9150    | 2     | 0.20%   |
| 46 | GVRH                | AS34756   | 2     | 0.20%   |
| 47 | UNILOGICNET         | AS28788   | 2     | 0.20%   |
| 48 | INTERNLNET          | AS20507   | 2     | 0.20%   |
| 49 | NOB                 | AS20969   | 2     | 0.20%   |
| 50 | VIRTU | AS16243 | 2 | 0.20% |
| 51 | QINIP | AS8608 | 1 | 0.10% |
| 52 | WERITECH | AS199139 | 1 | 0.10% |
| 53 | UNISCAPEB | AS201975 | 1 | 0.10% |
| 54 | BLACKGATE | AS201290 | 1 | 0.10% |
| 55 | undefined | AS200130 | 1 | 0.10% |
| 56 | SOURCEXS | AS56510 | 1 | 0.10% |
| 57 | STEPCO | AS57146 | 1 | 0.10% |
| 58 | GO-TREX | AS199752 | 1 | 0.10% |
| 59 | EXCL | AS60479 | 1 | 0.10% |
| 60 | INFRACOM | AS8587 | 1 | 0.10% |
| 61 | XRCSERVICES | AS61060 | 1 | 0.10% |
| 62 | KPNM- | AS1134 | 1 | 0.10% |
| 63 | IPVN01 | AS198089 | 1 | 0.10% |
| 64 | MICROSOFT-CORP-MSN-BLOCK | AS8075 | 1 | 0.10% |
| 65 | PETIT | AS197156 | 1 | 0.10% |
| 66 | MICAIP | AS203037 | 1 | 0.10% |
| 67 | WORLDSTREAM | AS49981 | 1 | 0.10% |
| 68 | DATAWEB | AS35332 | 1 | 0.10% |
| 69 | FUNDAMENTS | AS20559 | 1 | 0.10% |
| 70 | UTWENTE | AS1133 | 1 | 0.10% |
| 71 | NGI | AS35612 | 1 | 0.10% |
| 72 | YCC | AS31499 | 1 | 0.10% |
| 73 | DSD | AS29462 | 1 | 0.10% |
| 74 | GLOBAL-E | AS39591 | 1 | 0.10% |
| 75 | BUSINESSCONNECT | AS15693 | 1 | 0.10% |
| 76 | ATGS-MMD | AS2686 | 1 | 0.10% |
| 77 | EQUEST | AS42707 | 1 | 0.10% |
| 78 | DT-IT | AS42812 | 1 | 0.10% |
| 79 | NFORCE | AS43350 | 1 | 0.10% |
| 80 | KABELTEX | AS43995 | 1 | 0.10% |
| 81 | AMS-MARLINK-MSS | AS44933 | 1 | 0.10% |
| 82 | ISICONNEXION | AS44953 | 1 | 0.10% |
| 83 | LIBERNET | AS206389 | 1 | 0.10% |
| 84 | DCN | AS48812 | 1 | 0.10% |
| 85 | PELICAN-ICT | AS35705 | 1 | 0.10% |

*<This page was intentionally left blank.>*